\documentclass[12pt]{article}
\newcommand\citep[1]{\cite{#1}}
\newcommand\acknowledgments[1]{{\bf Acknowledgments}}

\usepackage{graphicx}
\usepackage{amsmath}

\newcommand{\be}{\begin{equation}}
\newcommand{\ee}{\end{equation}}
\newcommand\eqn[1]{$$#1$$}
\newcommand\eqnn[2]{
	\begin{equation}#2\label{#1}
	\end{equation}}
\newcommand\elnn[2]{
	\begin{eqnarray}
	#2\label{#1}
	\end{eqnarray}}
\newcommand\elnnsingle[2]{
	\begin{align}
	\begin{split}
	#2\label{#1}
	\end{split}
	\end{align}
}
\newcommand\elnnn[1]{
	\begin{eqnarray}
	#1
	\end{eqnarray}}

\begin{document}

\title{Galactic Dynamics via General Relativity: A Compilation and New Developments}
\author{F. I. Cooperstock and S. Tieu\\
	Department of Physics and Astronomy,
	University of Victoria\\
	P.O. Box 3055, Victoria, B.C. V8W 3P6 (Canada)\\
	{\small Email: cooperst@uvic.ca, stieu@uvic.ca} }

\maketitle

\abstract{
We consider the consequences of applying general relativity to the 
description of the dynamics of a galaxy, given the observed flattened 
rotation curves. The galaxy is modeled as a stationary axially 
symmetric pressure-free fluid. In spite of the weak gravitational 
field and the non-relativistic source velocities, the mathematical 
system is still seen to be non-linear.  It is shown that the rotation 
curves for various galaxies as examples are 
consistent with the mass density distributions of the visible matter 
within essentially flattened disks. This obviates the need for a massive 
halo of exotic dark matter. We determine that the mass density for 
the luminous threshold as tracked in the radial direction is 
$10^{-21.75}$ kg$\cdot$m$^{-3}$ for these galaxies and conjecture 
that this will be the case for other galaxies yet to be analyzed.  We 
present a velocity dispersion test to determine the extent, if of any 
significance, of matter that may lie beyond the visible/HI region. 
Various comments and criticisms from colleagues are addressed.
}


\section{Introduction}
This paper presents a unified amalgam of our work to date in 
\cite{CT}, \cite{CT2}, \cite{CT3} and the expansion of the Letter 
\cite{CT4} as well as further progress in the application of general 
relativity to galactic dynamics.  The motivation for the work stems 
from the need to acount for the observed essentially flat 
velocity rotation curves for galaxies. This has been a central issue 
in astrophysics. There has been much speculation over the question of 
the nature of the dark matter that is believed to be required for the 
consistency of the observations of higher than expected stellar 
velocities with Newtonian gravitational theory. 
While various researchers are now turning to gravitational lensing in the search for evidence for dark matter, probably the majority of researchers regard the flat galactic rotation curves as the key indicators.
Clearly the issue is 
of paramount importance given that the dark matter is said to 
comprise the dominant constituent of an extended galactic mass by 
multiple factors \cite{BT}.  The dark matter enigma has served as a 
spur for particle theorists to devise acceptable candidates for its 
constitution. While physicists and astrophysicists have pondered over 
the problem, other researchers have devised new theories of gravity to 
account for the observations
(see for example \cite{bek1,bek3,bek4,bek5}). However 
the latter approaches, imaginative as they may be, have met with 
understandable skepticism, having been devised solely for the purpose 
of the task at hand. To our knowledge, general relativity has never 
previously  been proposed as a possible means of accounting for the 
flat rotation curves without invoking large stores of dark matter. However, 
general relativity remains the preferred theory of gravity with 
Newtonian theory as its limit where appropriate. General relativity 
has been successful in every test that it has encountered, going 
beyond Newtonian theory where required. Therefore, should it actually 
transcend Newtonian gravity in resolving the problem at hand, it 
would greatly alter the understanding of some basic aspects in 
physics. In what follows, we will set out to show that this is the 
case.

It is understandable that the conventional gravity approach has 
focused upon Newtonian theory in the study of galactic dynamics as 
the galactic field is weak (apart from the deep core regions where 
black holes are said to reside, at least in some galaxies) and the 
motions are non-relativistic ($v \ll{c}$). It was this approach that 
led to the inconsistency between the theoretical Newtonian-based 
predictions and the observations of the visible sources alone. To 
reconcile the theory with the observations, researchers subsequently 
concluded that to realize the observed motions, dark matter must be 
present around galaxies in vast massive halos that constitute the 
great bulk of the extended galactic masses\footnote{
	See however \cite{keet} who argues for a much less massive
	halo based upon gravitational lensing data.}.  While this might
at first sight appear to be a relatively simple cure to the problem 
of motion, these massive halos cannot be identified with any known 
form of matter, hence our use of the adjective ``exotic'' to describe 
this presumed matter. However, in dismissing general relativity in 
favor of Newtonian gravitational theory for the study of galactic 
dynamics, insufficient attention has been paid to the fact that the 
stars that compose the galaxies are essentially in motion under 
gravity alone (``gravitationally bound''). It has been known since 
the time of Eddington that the gravitationally bound problem in 
general relativity is an intrinsically non-linear problem \footnote{
	It is to be noted that after nearly a century of research,
	there still does not exist a closed-form solution of the
	two-body problem in general relativity.
}
even when the conditions are such that the field is weak and the 
motions are non-relativistic, at least in the time-dependent case. 
\textit{Most significantly, we have found that under these 
conditions, the general relativistic analysis of the problem is also 
non-linear for the stationary (non-time-dependent) case at hand.} 
Thus the intrinsically linear Newtonian-based approach used to this 
point has been inadequate for the description of the galactic 
dynamics and Einstein's general relativity must be brought into the 
analysis within the framework of established gravitational 
theory\footnote{
	Actually within the framework of Newtonian theory, it is 
	possible to define an ``effective'' potential (see for 
	example \cite{BT} page 136) to incorporate the centrifugal 
	acceleration in a rotating coordinate system with a given 
	angular velocity. Since this contains the square of the 
	angular velocity of the rotating frame, there is already 
	the hint of non-linearity present.  However, in what 
	follows in general relativity, we will see the 
	non-linearity related to the angular velocity as a 
	\textit{variable} function. Moreover, for a system in 
	rotation, this non-linearity cannot be removed globally.
}. This is an essential departure from conventional thinking on the 
subject and it leads to major consequences as we discuss in what 
follows. We will demonstrate that via general relativity, what we will refer to as 
``generating" potentials producing the observed flattened  rotation 
curves can be linked to the mass density distributions of the 
essentially flattened disks, obviating any necessity for dominant massive 
exotic dark matter halos in the total extended galactic composition.

We will also present the indicator that the threshold for luminosity, 
as we probe in the radial direction, occurs at a density of 
$10^{-21.75}$ kg$\cdot$m$^{-3}$. 

Since our initial posting \cite{CT}, many colleagues have offered 
their comments and criticisms which we address in Section 4. Some of 
these issues have already been discussed in \cite{CT2}, \cite{CT3} 
and \cite{CT4}. In this paper, we expand upon our discussions of the 
various papers. We also focus upon a new observational discriminator 
for assessing the degree, if any, of external matter that may lie 
beyond the visible/HI regions.

\section{The Model Galaxy}

Within the context of Newtonian theory, Mestel \cite{mes} considered 
a special rotating disk with surface density inversely proportional 
to radius. Using a disk potential with Bessel functions that we will 
also use in what follows but in quite a different manner, he found 
that it leads to an absolutely flat galactic rotation velocity 
curve.\footnote{
	This is also the case for the MOND
	\cite{bek1,bek3,bek4} model.
}
Interestingly, the gradient of the potential in this, as in all 
Newtonian treatments, relates to acceleration whereas in the general 
relativistic treatment, we will show that the gradient of a 
generating potential gives the stellar tangential velocity (\ref{Eq11}). 

When we consider the complexity of the detailed structure of a spiral 
galaxy With its arms and irregular density variations, it becomes 
clear that the modeling within the context of the complicated theory 
of general relativity must entail some simplifications. As long as 
the essence of the structure is captured, these simplifications are 
justified and valuable information can be gleaned. 
Thus, in terms of its essential characteristics, we consider a 
uniformly rotating fluid without pressure and symmetric about its 
axis of rotation. We do so within the context of general relativity. 
The stationary axially symmetric metric can be described in 
generality in the form
\eqnn{Eq1}{
	ds^2 =	-e^{\nu-w}( udz^2+dr^2)
		-r^2 e^{-w} d\phi^2+e^w(cdt-Nd\phi)^2
}
where $u$, $\nu$, $w$ and $N$ are functions of cylindrical polar 
coordinates $r$, $z$. It is easy to show that to the order required, 
$u$ can be taken to be unity. \footnote{
	Retaining terms of non-zero order in $G$ for $u$ induces
	terms of order $G^n$ with $n>1$ in the field equations.
}
It is most simple to work in the frame that is co-moving with the
matter,
\eqnn{Eq2}{
	U^i = {\delta}_0^i
}
where $U^i$ is the four-velocity\footnote{
	This is reminiscent of the standard approach that is
	followed for FRW cosmologies. However, the FRW spacetimes
	are homogeneous and they are not stationary.
}.
This was done in the pioneering paper by van Stockum \cite{vs} who 
set $w=0$ from the outset\footnote{
	Interestingly, the geodesic equations imply that
	$w=constant$ (which can be taken to be zero as in
	\cite{vs}) even for the \textit{exact} Einstein field
	equations as studied in \cite{vs}. In fact the
	requirement that $w=0$ can be seen directly using
	(\ref{Eq2}) and the metric equation $g_{ik}U^iU^k=1$
	\cite{CT2}.
}. As in \cite{Bonnor}, and following \cite{Bardeen}, we perform
a purely \textit{local} ($r,z$ held fixed at each point when
taking differentials) transformation
\footnote{
Note that this local transformation is used only to deduce the connection between $N$ and $\omega$ (and hence $V$). All subsequent work continues in the original unbarred co-moving frame.
} 
\eqnn{Eq3}{
	\bar{\phi} = \phi + \omega(r,z)\,t
}
that locally diagonalizes the metric. In this manner, we are able to 
deduce the local angular velocity $\omega$ and the tangential 
velocity $V$ as
\elnn{Eq4}{
	&\omega
	= \frac{Nce^w}{r^2e^{-w}-N^2e^w}
	\approx \frac{Nc}{r^2} \label{Eq4top}\\
      & V      
      =\omega r
}
with the approximate value applicable for the weak fields under 
consideration. The Einstein field equations to order $G^1$
with $w$ retained for later comparison, are\footnote{
	This is a loose notation favored by many relativists but
	adequate for our purposes here as a smallness parameter.
}
\elnnsingle{Eq5top}{
	2r\nu_r+ N_r^2-N_z^2 =0, \\
	r\nu_z +N_r N_z =0,  \\
	N_r^2 + N_z^2 +2r^2(\nu_{rr}+\nu_{zz}) =0, \\
	N_{rr} +N_{zz} - \frac{N_r}{r}=0,
}
\elnn{Eq5}{
	\left(w_{rr} +w_{zz} +\frac{w_r}{r}\right)
	+ \frac{3}{4}r^{-2} (N_r^2 + N_z^2)& \nonumber \\
	+ \frac{N}{r^2}\left(N_{rr} +N_{zz} -\frac{N_r}{r}\right)
	- \frac{1}{2}(\nu_{rr}+\nu_{zz})
	&= 8{\pi}G\rho/ c^2
}
where $G$ is the gravitational constant and $\rho$ is the mass 
density. Subscripts denote partial differentiation with respect to 
the indicated variable. These equations are easily combined to yield
\eqnn{Eq5a}{
	\nabla^2 w +\frac{N_r^2+N_z^2}{r^2}=\frac{8\pi G\rho}{c^2}
}
where the first term is the flat-space Laplacian in cylindrical polar coordinates
\eqnn{Eq5b}{
	\nabla^2 w \equiv w_{rr} + w_{zz} + \frac{w_r}{r}
}
and $\nu$ would be determined by quadratures.

With the freely gravitating constraint and the requirement that $w=0$ 
arising from the choice of co-moving coordinates, the field equations 
for $N$ and $\rho$ in this globally dust distribution are reduced to 
\footnote{
	Note that with the minus sign in (\ref{Eq9a}), $N$ does not
	satisfy the Laplace equation.
}
\elnnn{
	N_{rr} + N_{zz} - \frac{N_r}{r} =0
	\label{Eq9a} \\
	\frac{N_r^2 + N_z^2}{r^2} = \frac{8{\pi}G\rho}{c^2}.
	\label{Eq9b}
}
Note that from both the field equation for $\rho$ and the expression 
for $\omega$ that $N$ is of order $G^{1/2}$. The non-linearity of 
the galactic dynamical problem is manifest through the
non-linear relation\footnote{
	While we have eliminated $w$ either by using the geodesic
	equations to get (\ref{Eq9b}) or by the metric equation and
	the choice of co-moving coordinates, $N$ cannot be eliminated
	and hence non-linearity is intrinsic to the study of the
	galactic dynamics.
}
between the functions $\rho$ and $N$. Rotation under freely 
gravitating motion is the key here. By contrast, for 
time-independence in the non-rotating problem, there must be pressure 
present to maintain a static configuration (hence altering the right 
hand side of (\ref{Eq5top})), $N$ vanishes for vanishing $\omega$ and 
$\nabla^2 w$ is non-zero yielding the familiar Poisson equation of 
Newtonian gravity. In the present case, it is the \textit{rotation} 
via the function $N$ that connects directly to the density and the 
now non-linear equation is in sharp contrast to the linear Poisson 
equation. 

Interestingly, (\ref{Eq9a}) can be expressed as
\eqnn{Eq10}{
	\nabla^2\Phi =0
}
where
\eqnn{Eq10a}{
	\Phi \equiv \int\frac{N}{r}dr
}
and hence flat-space harmonic functions $\Phi$ are the generators of 
the axially symmetric stationary pressure-free weak fields that we
seek\footnote{
	In fact Winicour \cite{Winicour} has shown that all such sources,
	even when the fields are strong, are generated by such flat-space
	harmonic functions.
}. 
These are the generating potentials referred to earlier. It is to be noted that these generating potentials play a different role in general relativity than do the potentials of Newtonian gravitational theory even though both functions are harmonic. 
Using (\ref{Eq4}) and (\ref{Eq10a}), we have the expression for 
the tangential velocity of the distribution
\eqnn{Eq11}{
	V=c\frac{N}{r} =c\frac{\partial {\Phi}}{\partial{r}}
}

\section{Modeling the Observed Galactic Rotation Curves}

\begin{figure}
\begin{center}
\includegraphics[width=2.75in]{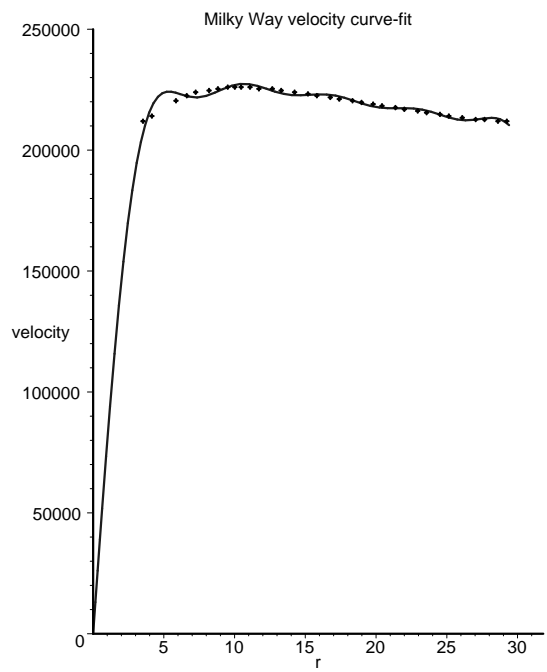}
\end{center}
\caption{
\label{fig:milkywayvelocity}
	Velocity curve-fit for the Milky Way
	in units of m/s vs Kpc.
}
\end{figure}

\begin{figure}
\begin{center}
\includegraphics[width=2.5in]{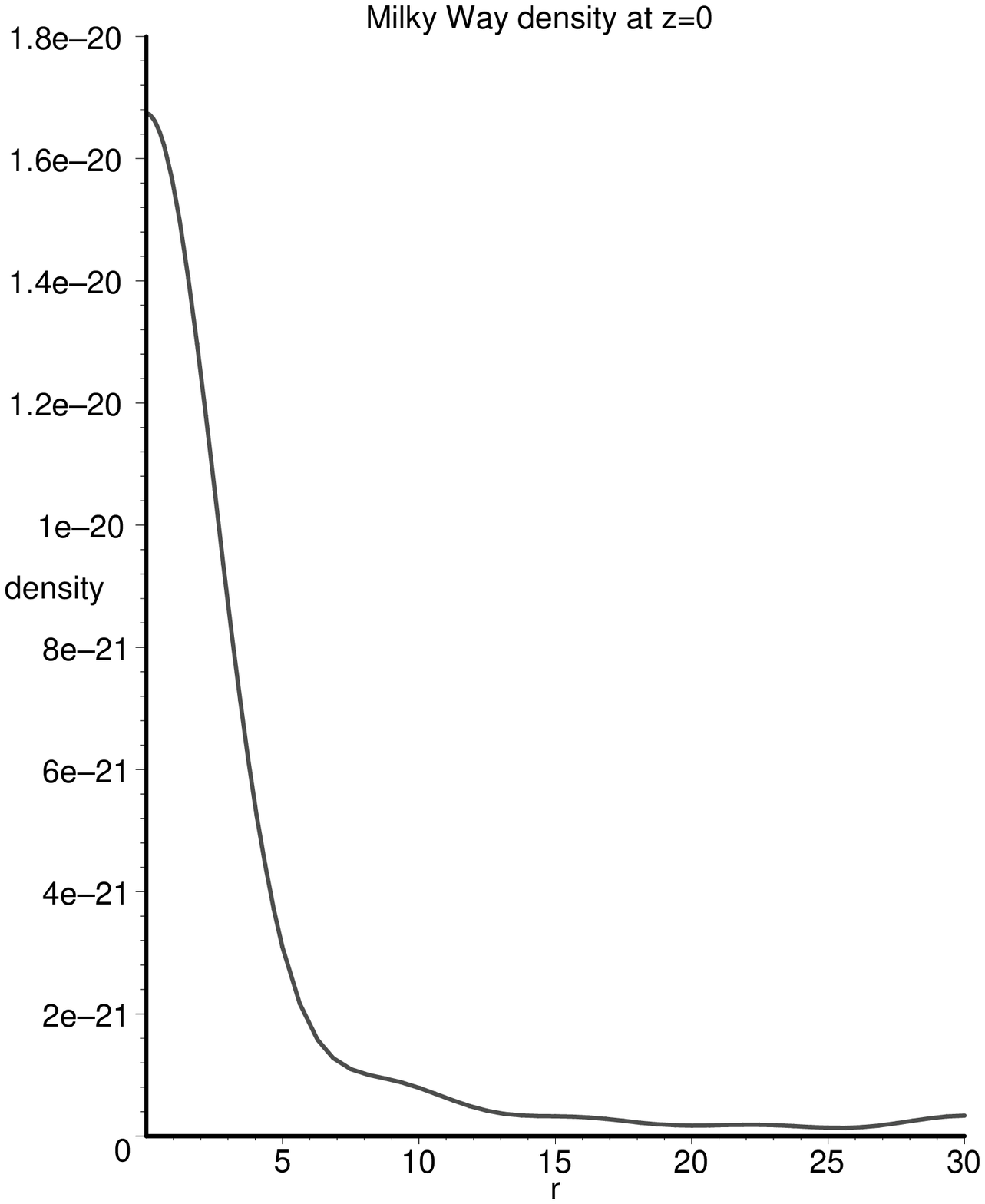}
\vskip 0.25in 
\includegraphics[width=2.5in]{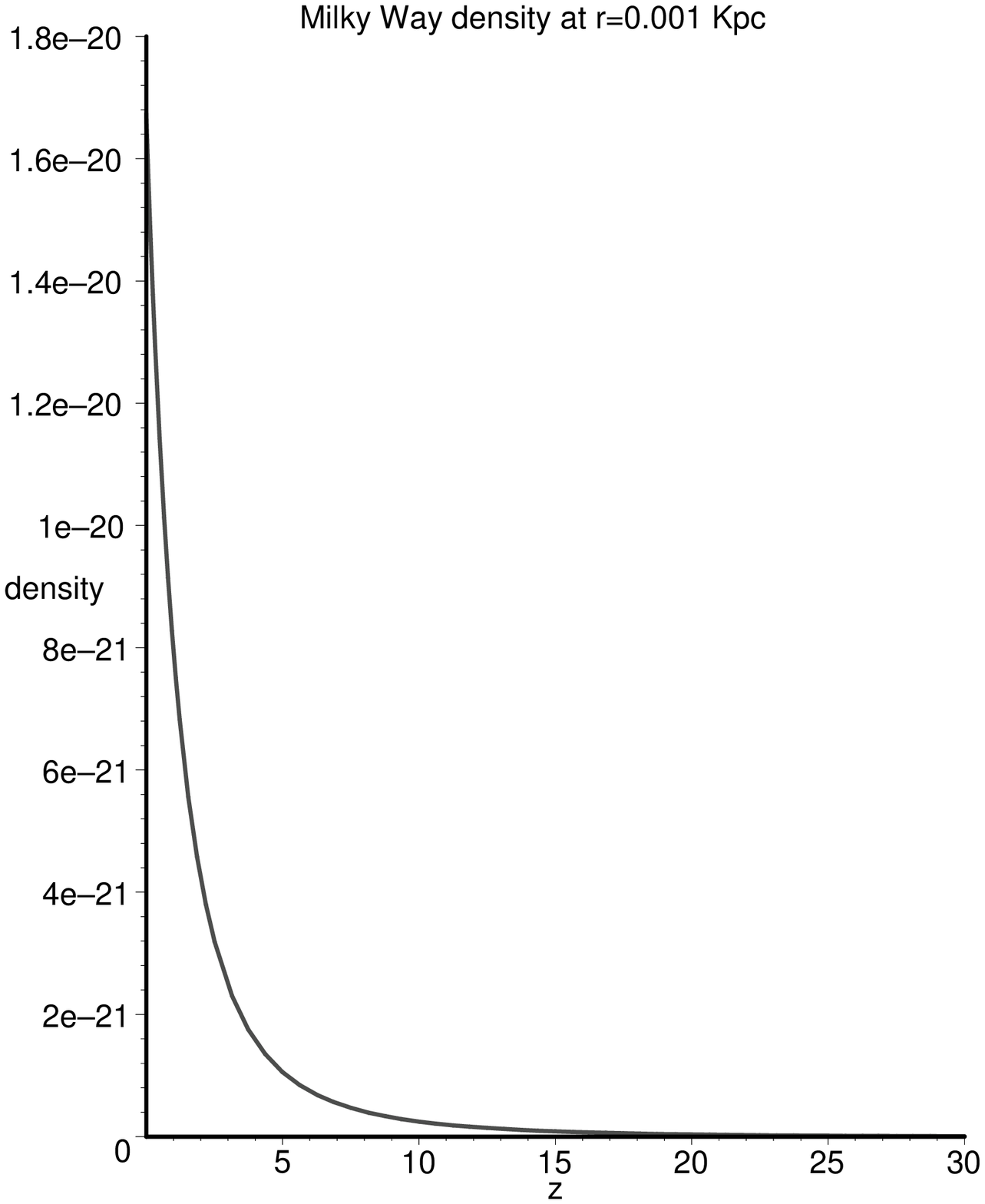}
\end{center}
\caption{
\label{fig:milkywaydensity}
	Derived density profiles in units of kg/m$^3$
	for  the Milky Way at (a) $z=0$
	and (b) $r=0.001$ Kpc.
}
\end{figure}

Since the field equation for $\rho$ is non-linear, the simpler way to 
proceed in galactic modeling is to first find the required generating 
potential $\Phi$ and from this, derive an appropriate function $N$ 
for the galaxy that is being analyzed. With $N$ found, (\ref{Eq9b}) 
yields the density distribution. If this is in accord with 
observations, the efficacy of the approach is established. This is in 
the reverse order of the standard approach to solving gravitational 
problems but it is most efficient in this formalism because of the 
existence of one linear field equation.

\begin{figure}
\begin{center}
\includegraphics[width=3in]{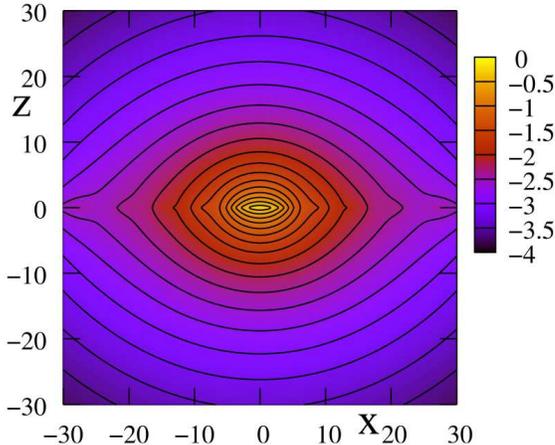}
\end{center}
\caption{
\label{fig:contour}
Cross-sectional density contour plot for the Milky Way model.
}
\end{figure}

Every galaxy is different and each requires its own composing 
elements to build the generating potential. In cylindrical polar 
coordinates, separation of variables yields the following solution 
for $\Phi$ in (\ref{Eq10}):
\eqnn{Eq12}{
	\Phi = Ce^{-k\mid z \mid}J_0(kr)
}
where $J_0$ is the Bessel function $m=0$ of Bessel $J_m(kr)$ and $C$
is an arbitrary constant. \footnote{
	See for example \cite{AW}. With this form of solution, the 
	absolute value of $z$ must be used to provide the proper 
	reflection of the distribution for negative $z$. While this 
	produces a discontinuity in $N_z$ at $z=0$, it is important 
	to note that in the problem at hand, this discontinuity is 
	consistent with the general case of having a density 
	gradient discontinuity at the plane of reflection symmetry. 
	This point is discussed further in the text.
} 
We use the linearity of (\ref{Eq10}) to express the general
solution of this form as a linear superposition
\eqnn{Eq13}{
	\Phi = \sum_{n}C_ne^{-k_n |z|}J_0(k_nr)
}
with $n$ chosen appropriately for the desired level of accuracy.
From (\ref{Eq13}) and (\ref{Eq11}), the tangential
velocity \footnote{
	$dJ_0(x)/dx= - J_1(x)$ from \cite{ford}.
} is
\eqnn{Eq14}{
	V= -c\sum_{n} k_n C_n e^{-k_n |z|}J_1(k_nr)
}
With the $k_n$ chosen so that the $J_0(k_nr)$ terms are orthogonal
\footnote{
	Just as the $\sin kx$ functions are orthogonal for integer 
	$k$, the Bessel functions $J_0(kr)$ have their own 
	orthogonality relation: $\int_{0}^{1} J_0(k_nr)J_0(k_mr)rdr 
	\propto \delta_{mn}$ where $k_n$ are the zeros of $J_0$ at 
	the limits of integration. This orthogonality condition is 
	on $\Phi$ rather than on $V$ because the differential 
	equation dictates the integral condition.
}
to each other, we have found that only 10 functions with parameters 
$C_n$, $n\in\{1\dots 10\}$ suffice to provide an excellent
fit\footnote{
	It should be noted that unlike typical velocity curve fits 
	that allow arbitrary velocity functions, our curve fits are 
	constrained by the demand that they be created from 
	derivatives of harmonic functions.
}
to the velocity curve for the Milky Way. The details are
provided in the Appendix and the curve fit is shown in
Figure \ref{fig:milkywayvelocity}.
\footnote{
	Note that the $J_1(x)$ Bessel functions are $0$ at $x=0$ 
	and oscillate with decreasing amplitude, falling as 
	$1/\sqrt{x}$ asymptotically \cite{ford}.  
	However, this feature alone does not assure a realistic 
	fall-off of matter. This issue is addressed in Section 4. 
	Also, the present curves drop as $r$ approaches $0$.  This 
	is in contrast to the Mestel \cite{mes} and MOND
	\cite{bek1,bek3,bek4} curves that are flat
	everywhere. 
} From (\ref{Eq11}) and (\ref{Eq14}), the $N$ function is determined 
in detail and from (\ref{Eq9b}), the density distribution follows.  
This is shown in Figure \ref{fig:milkywaydensity} as a function of 
$r$ at $z=0$ as well as a function of $z$ at $r=0.001$ Kpc. We see 
that the distribution is an essentially flattened disk with good 
correlation with the observed overall averaged density data for the 
Milky Way (see Figure \ref{fig:contour}). 
\footnote{
In Figure \ref{fig:contour}, we see that in the cross-sectional density plot, the equidensity contours are approximately elliptical around the visible galactic region.
} 
The integrated mass is found to be $21 \times 
10^{10}M_\odot$ which is at the lower end of the estimated mass range 
of $20 \times 10^{10}M_\odot$ to $60 \times 10^{10}M_\odot$ as 
established by various researchers.  It is to be noted that the 
approximation scheme would break down in the region of the galactic 
core should the core harbor a black hole or even a naked singularity 
(see e.g.  \cite{coop}). \textit{Most significantly, our correlation 
of the flat velocity curve is achieved with disk mass of an order of 
magnitude smaller than the envisaged halo mass of exotic dark 
matter.} \footnote{
	See e.g.\cite{clew1,clew2} for proposed values of extended halo 
	masses. See also our discussions in Sections 4 and 5 of
	the added mass in extended dust distributions. 
}

General relativity does not distinguish between the luminous and 
non-luminous contributions. The deduced $\rho$ density distribution 
is derived from the totality of the two. Any substantial amount of 
non-luminous matter (i.e.  \textit{conventional} non-exotic dark 
matter) would necessarily lie in the flattened region 
relatively close to $z=0$ because this is the region of significant 
$\rho$ and would be due to dead stars, planets, neutron stars and 
other normal non-luminous  baryonic matter debris. Each term within 
the series has $z$-dependence of the form $e^{-k_n|z|}$ which causes the steep density fall-off profile as shown in Figure 
\ref{fig:milkywaydensity}(b).  This fortifies the picture of a 
standard galactic essentially flattened disk-like shape as 
opposed to a halo sphere. From the evidence provided thus far by 
rotation curves \footnote{
	See Section 5 for a proposed observational test for
	assigning a measure of mass to the region extending
	beyond the visible/HI region.
}, there is no support for the widely accepted notion of the 
necessity for massive halos of exotic dark matter surrounding visible 
galactic disks: conventional gravitational theory, namely general 
relativity, can account for the observed flat galactic rotation 
curves linked to essentially flattened disks with no evident need for 
exotic dark matter.

\begin{figure}
\begin{center}
\includegraphics[width=2.75in]{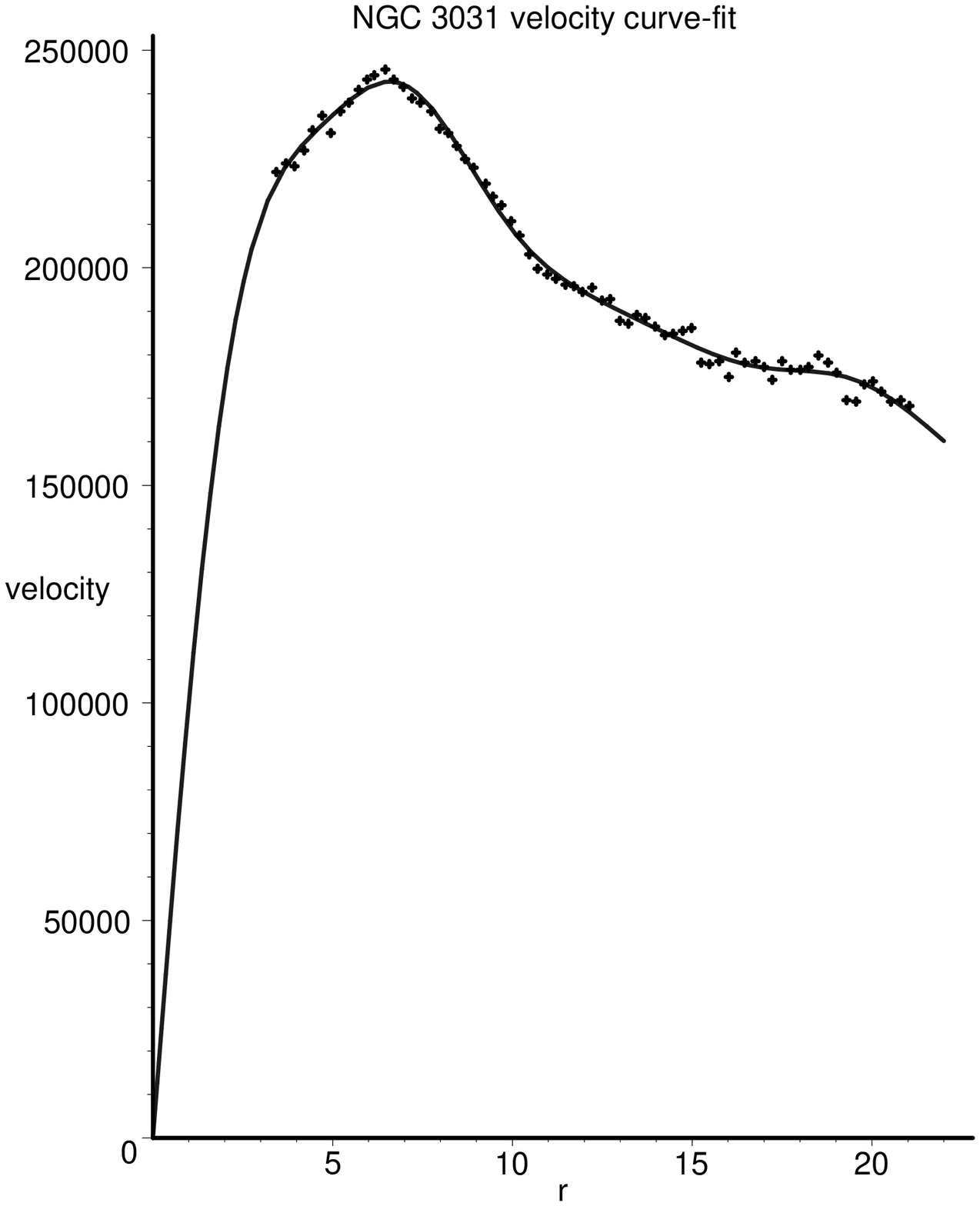}
\vskip 0.25in
\includegraphics[width=2.75in]{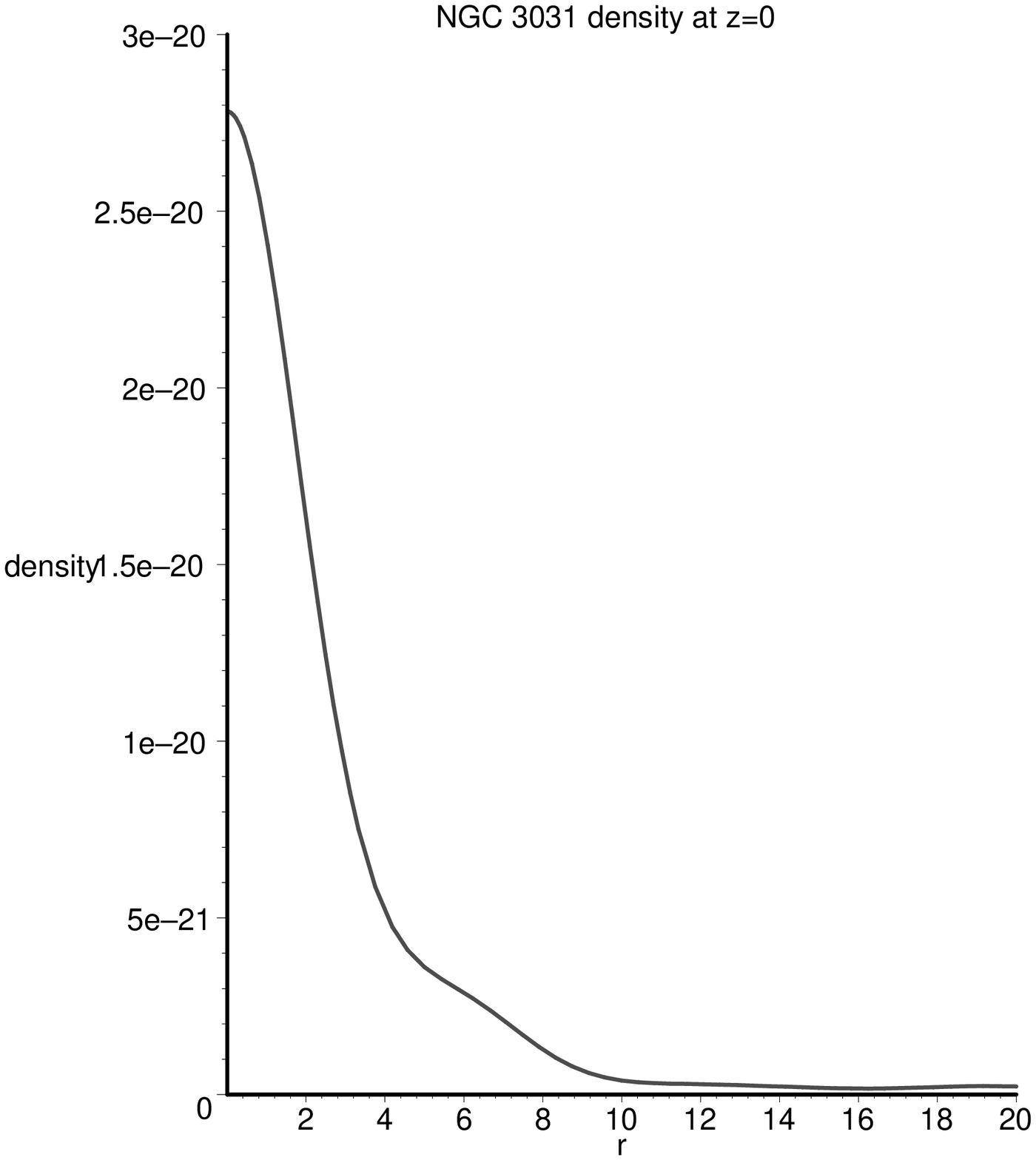}
\end{center}
\caption{
\label{fig:ngc3031}
Velocity curve-fit and derived density for NGC 3031
}
\end{figure}

\begin{figure}
\begin{center}
\includegraphics[width=2.75in]{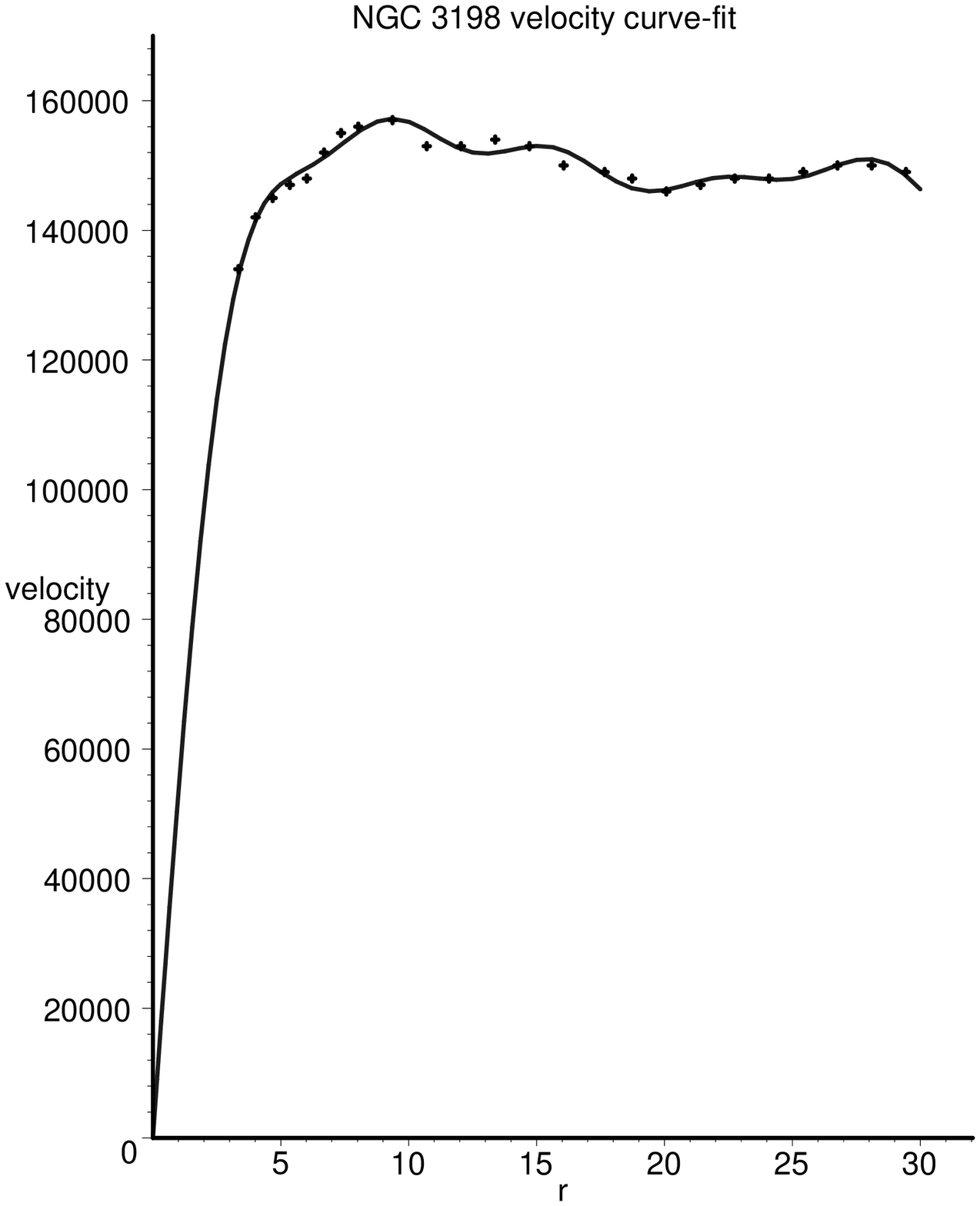}
\vskip 0.25in 
\includegraphics[width=2.75in]{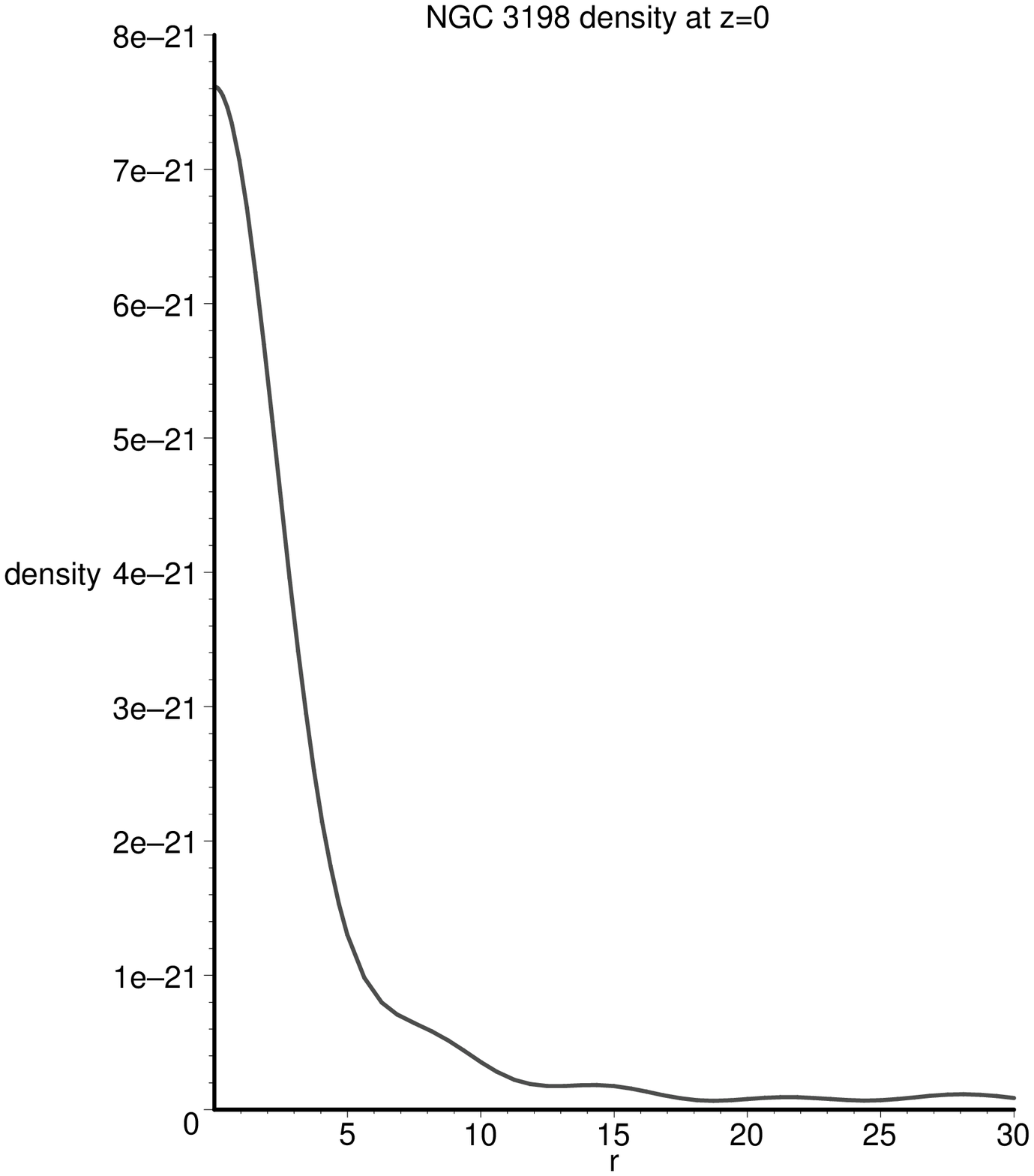}
\end{center}
\caption{
\label{fig:ngc3198}
Velocity curve-fit and
derived density for NGC 3198
}
\end{figure}

\begin{figure}
\begin{center}
\includegraphics[width=2.5in]{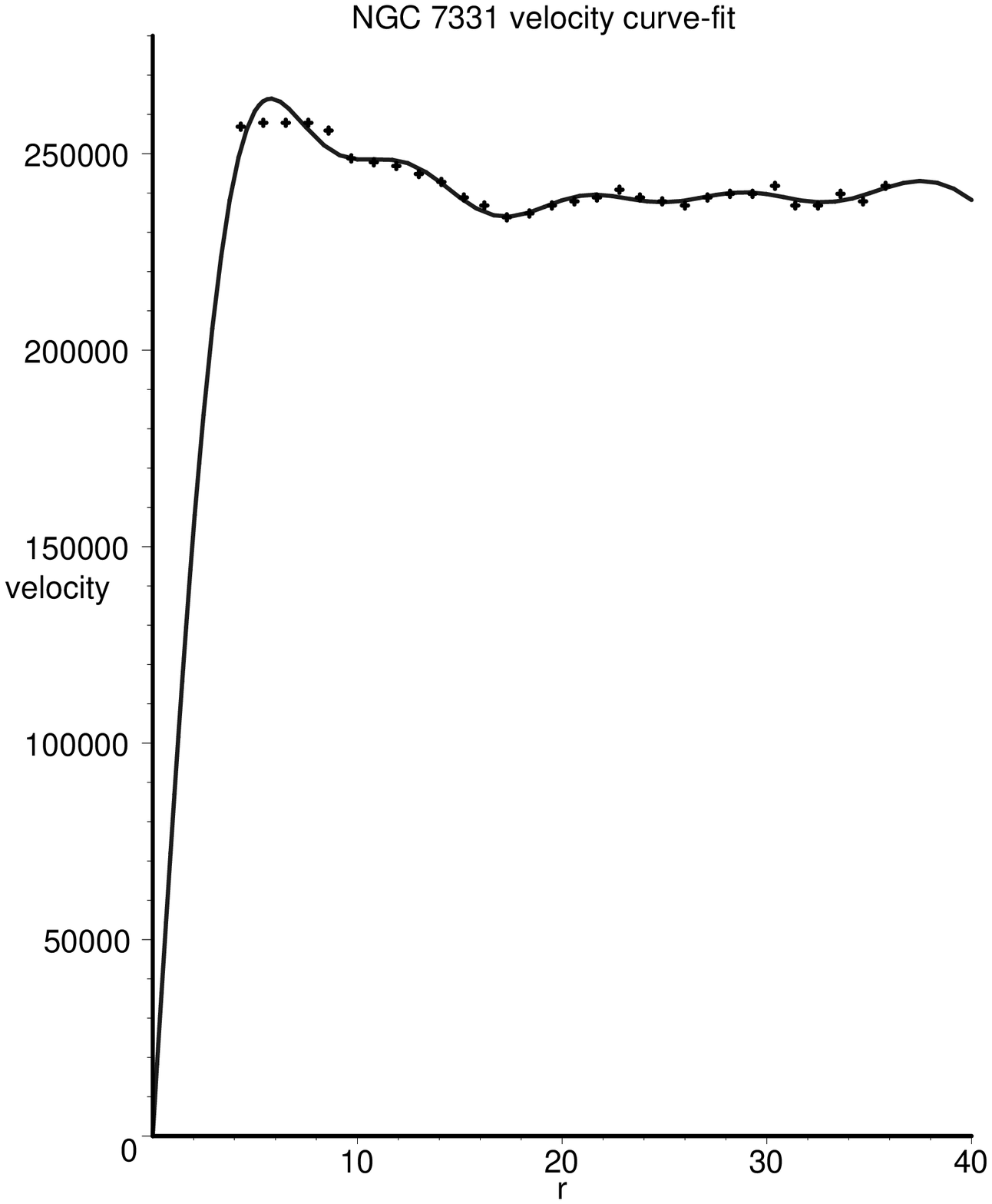}
\vskip 0.25in 
\includegraphics[width=2.5in]{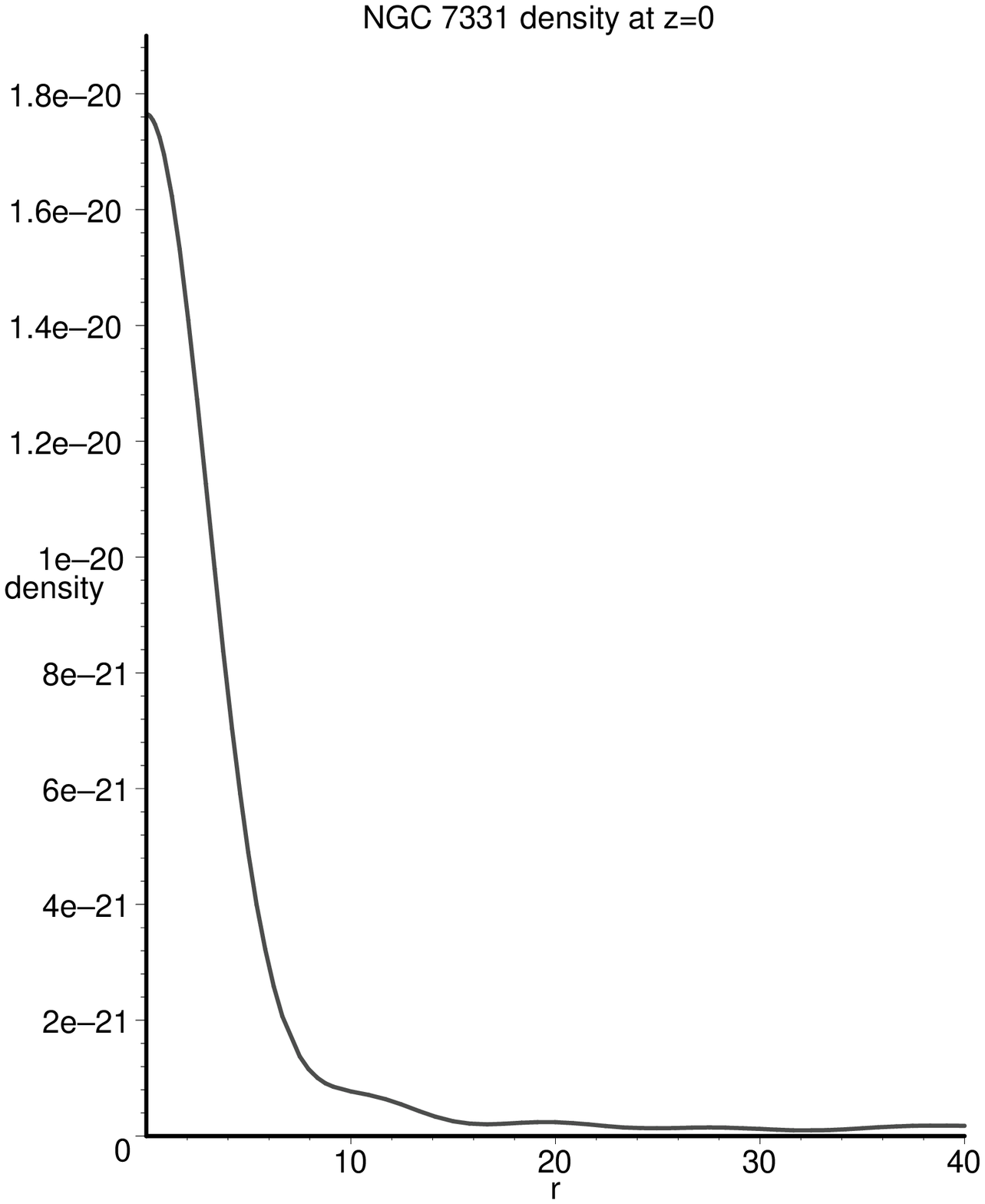}
\end{center}
\caption{
\label{fig:ngc7331}
Velocity curve-fit and derived density for NGC 7331
}
\end{figure}

We have also performed curve fits for the galaxies NGC 3031, NGC 3198 
and NGC 7331. The data are provided in the Appendix and the 
remarkably precise velocity curve fits are shown in Figures 
\ref{fig:ngc3031} to \ref{fig:ngc7331} where the density profiles are 
presented for $r$ at $z=0$. Again the picture is consistent with the 
observations and the mass is found to be $10.1 \times 10^{10}M_\odot$ 
for NGC 3198. This can be compared to the result from Milgrom's 
\cite{bek1,bek3,bek4} modified Newtonian dynamics of
$4.9 \times 10^{10}M_\odot$ and the value given through observations
(with Newtonian dynamics) by Kent \cite{Kent} of $15.1 \times 
10^{10}M_\odot$. While the visible light profile terminates at $r=14$ 
Kpc, the HI profile extends to 30 Kpc. If the density is integrated 
to 14 Kpc, it yields a mass-to-light ratio of $7\Upsilon_\odot$. 
However, integrating through the HI outer region to $r=30$ Kpc yields 
$14\Upsilon_\odot$ using data from \cite{alba}.

For NGC 7331, we calculate a mass of $26.0 \times 10^{10}M_\odot$. 
Kent \cite{Kent} finds a value of $43.3 \times 10^{10}M_\odot$. For 
NGC 3031, the mass is calculated to be $10.9\times 10^{10}M_\odot$ 
as compared to Kent's value of $13.3\times 10^{10}M_\odot$. Our 
masses are consistently lower than the masses projected by models 
invoking exotic dark matter halos and our distributions roughly tend 
to follow the contours of the optical disks.

\begin{figure}
\begin{center}
\includegraphics[width=2.75in]{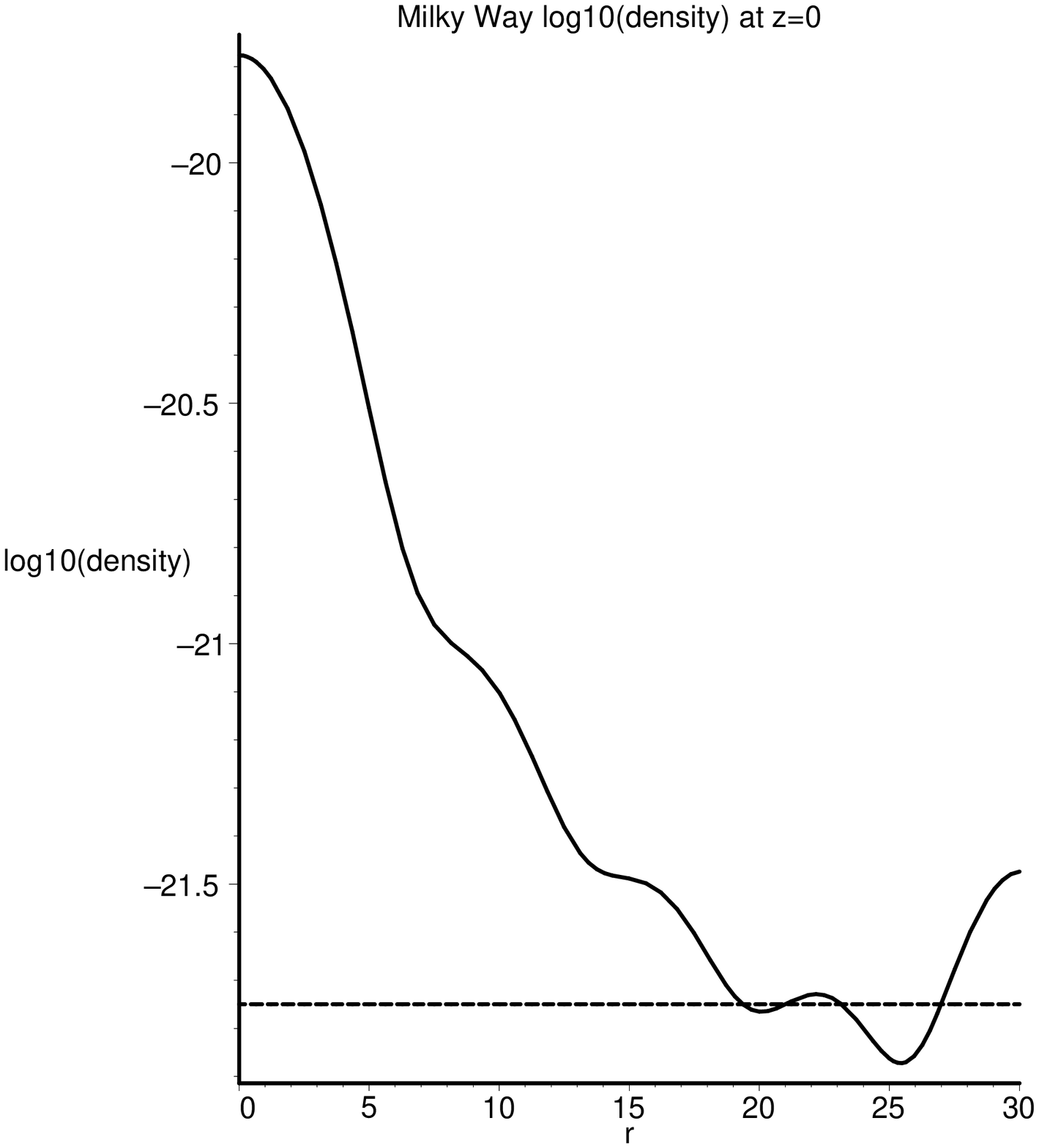}
\vskip 0.25in 
\includegraphics[width=2.75in]{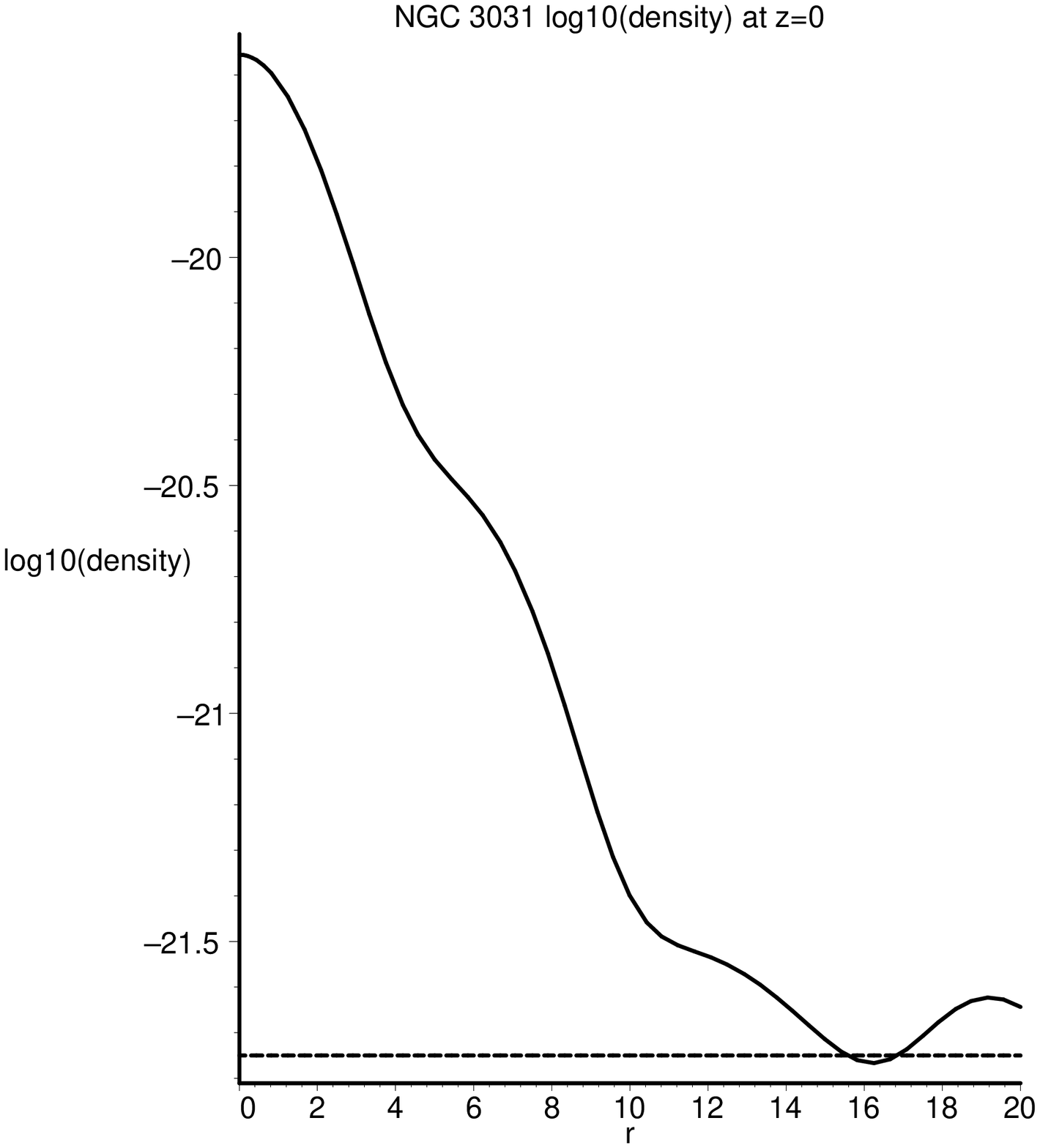}
\end{center}
\caption{
\label{fig:first2logdensity}
	Log graphs of density for (a) the Milky Way  and (b)
	NGC 3031 showing the density fall-off.
	The $-21.75$ dashed line provides a tool to predict
	the outer limits of visible matter. The fluctuations at
	the end are the result of limited curve-fitting terms.
}
\end{figure}

\begin{figure}
\begin{center}
\includegraphics[width=2.75in]{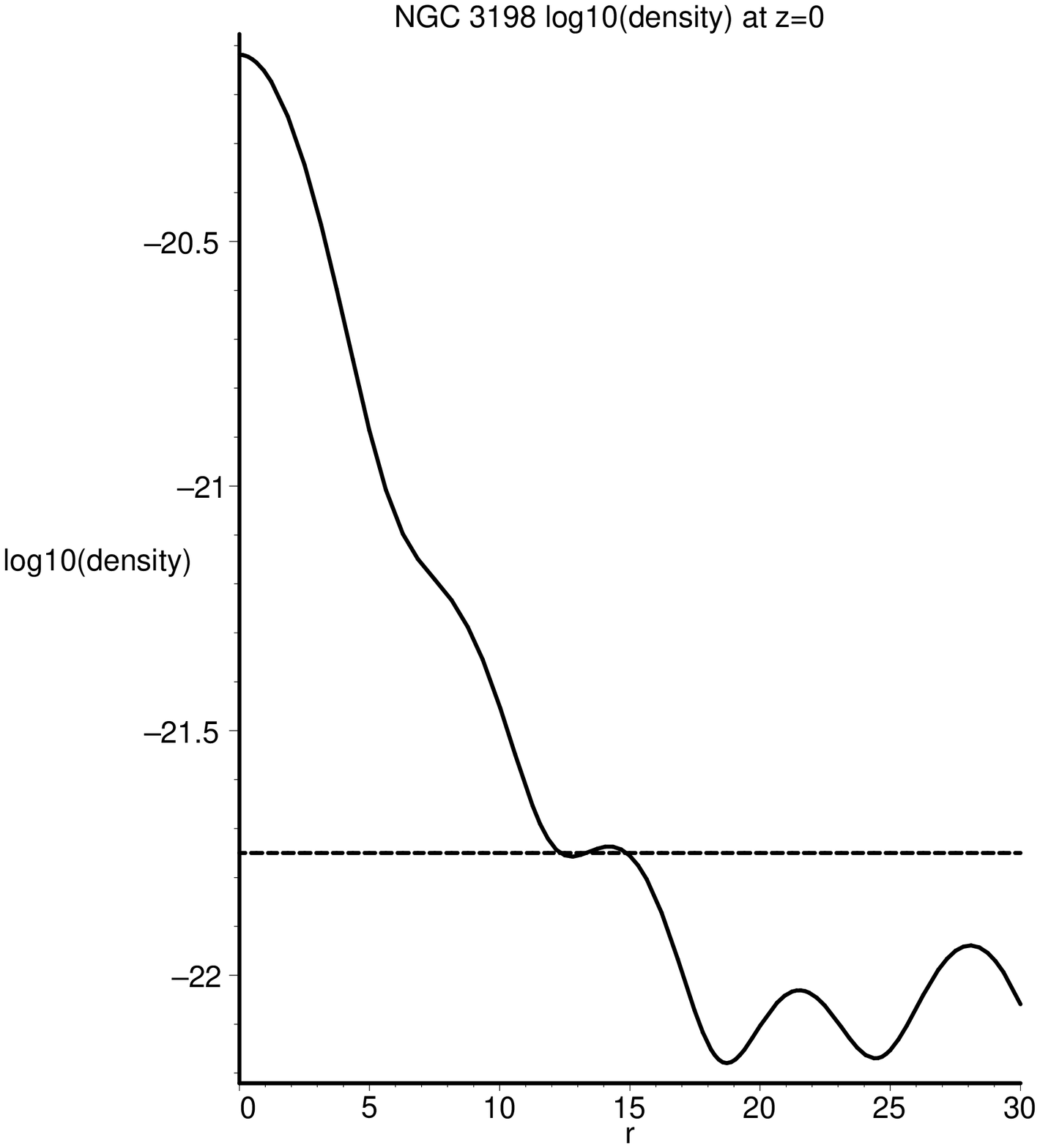}
\vskip 0.25in 
\includegraphics[width=2.75in]{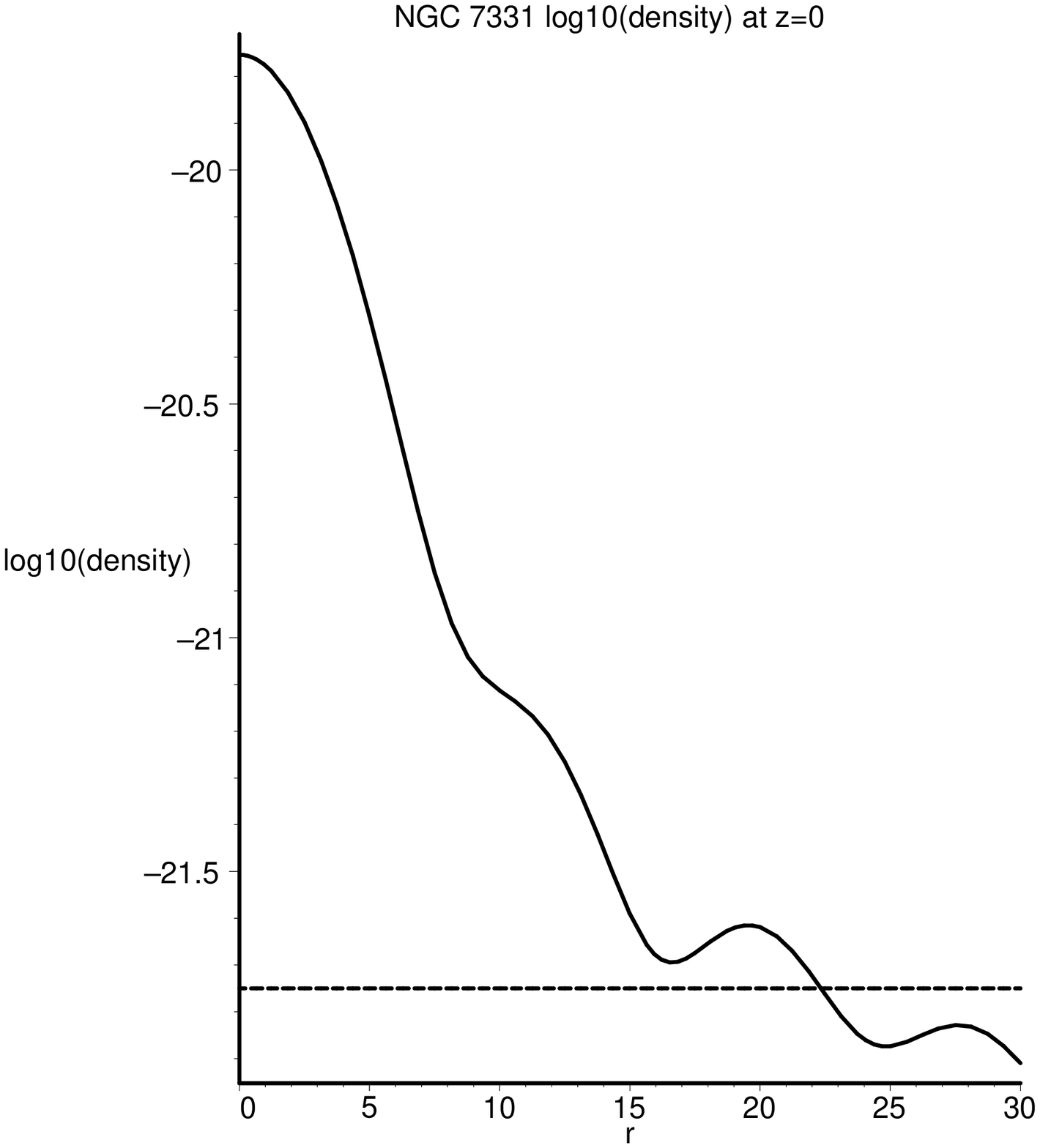}
\end{center}
\caption{
\label{fig:last2logdensity}
	Log graphs of density for (a) the NGC 3198 and (b)
	NGC 7331 showing the density fall-off.
	The $-21.75$ dashed line provides a tool to predict
	the limits of luminous matter. As before, there are
	fluctuations near the border.
}
\end{figure}

It is interesting to note that from the figures provided by Kent 
\cite{Kent} for optical intensity curves and our log density profiles 
for NGC 3031, NGC 3198 and NGC 7331, we determine that the threshold 
density for the onset of visible galactic light as we probe in the 
radial direction is at $10^{-21.75}$ kg$\cdot$m$^{-3}$ (Figure 
\ref{fig:first2logdensity} and Figure \ref{fig:last2logdensity}). It 
would be of interest to explore as many sources as possible to test 
the indicated hypothesis that this density is the universal optical 
luminosity threshold for galaxies as tracked in the radial direction. 
 Alternatively, should this hypothesis be further substantiated, the 
radius at which the optical luminosity fall-off occurs can be 
predicted for other sources using this special density parameter. The 
predicted optical luminosity fall-off for the Milky Way is at a 
radius of 19-21 Kpc based upon the density threshold indicator that 
we have determined.

Various authors attempt to incorporate the Tully- Fisher law 
\cite{tul} into their modified theories of gravity. General 
relativity can provide an equivalent albeit considerably more 
complicated relation but in integral form. From (\ref{Eq9a}) and 
(\ref{Eq11}), the radial gradient of the galactic mass can be 
expressed in terms of velocity as
\eqnn{Eq15}{
	M_r(r)
	=\frac{1}{2G}\int_{0}^{\infty}
	\left(r\left(V_r^2+V_z^2\right)
	+ \frac{V^2}{r} +2VV_r \right)\, dz
}
and a doubling has been used to account for the lower disk
contribution.

\section{Replies to Various Analyses}

Since the initial posting of our paper \cite{CT}, there has been an 
interesting variety of critical papers as well as two papers lending support to our work \cite{BG} \cite{Lus}. Some of the issues arising from these papers have been addressed
in our companion paper \cite{CT2}, in \cite{CT3} and in 
\cite{CT4}. For completeness, in what follows we will reply to all of 
the most relevant papers to date in this section as well as the 
frequently asked questions and comments proffered by our colleagues.

An issue first raised privately to us by some colleagues and later in 
\cite{korz} \cite{VL}  concerns the nature of the matter 
distribution. They have noted that given the existence of the 
discontinuity of $N_z$ that we had pointed to in \cite{CT}, a 
significant surface tensor $S_i^k$ can be constructed with a surface 
density component given by
\eqnn{Eq1r}{
	(8\pi G/c^2) S_t^t = \frac{N[N_z]}{2r^2} -\frac{[\nu_z]}{2}
}
to order $G^1$. The notation $[..]$ denotes the jump over a 
discontinuity of the given function, here at $z=0$. Using 
(\ref{Eq5top}), this becomes
\eqnn{Eq2r}{
	(8\pi G/c^2) S_t^t = \frac{N[N_z]}{2r^2} +\frac{N_r[N_z]}{2r}
}
It was claimed that this necessarily implied the existence of a singular 
\textit{physical} surface of mass in the galactic plane above and 
beyond the continuous mass distribution that we had found, thus 
rendering our model unphysical.

\begin{figure}
\begin{center}
\begin{tabular}{c c}
\begin{tabular}{c}
\includegraphics[width=1in]{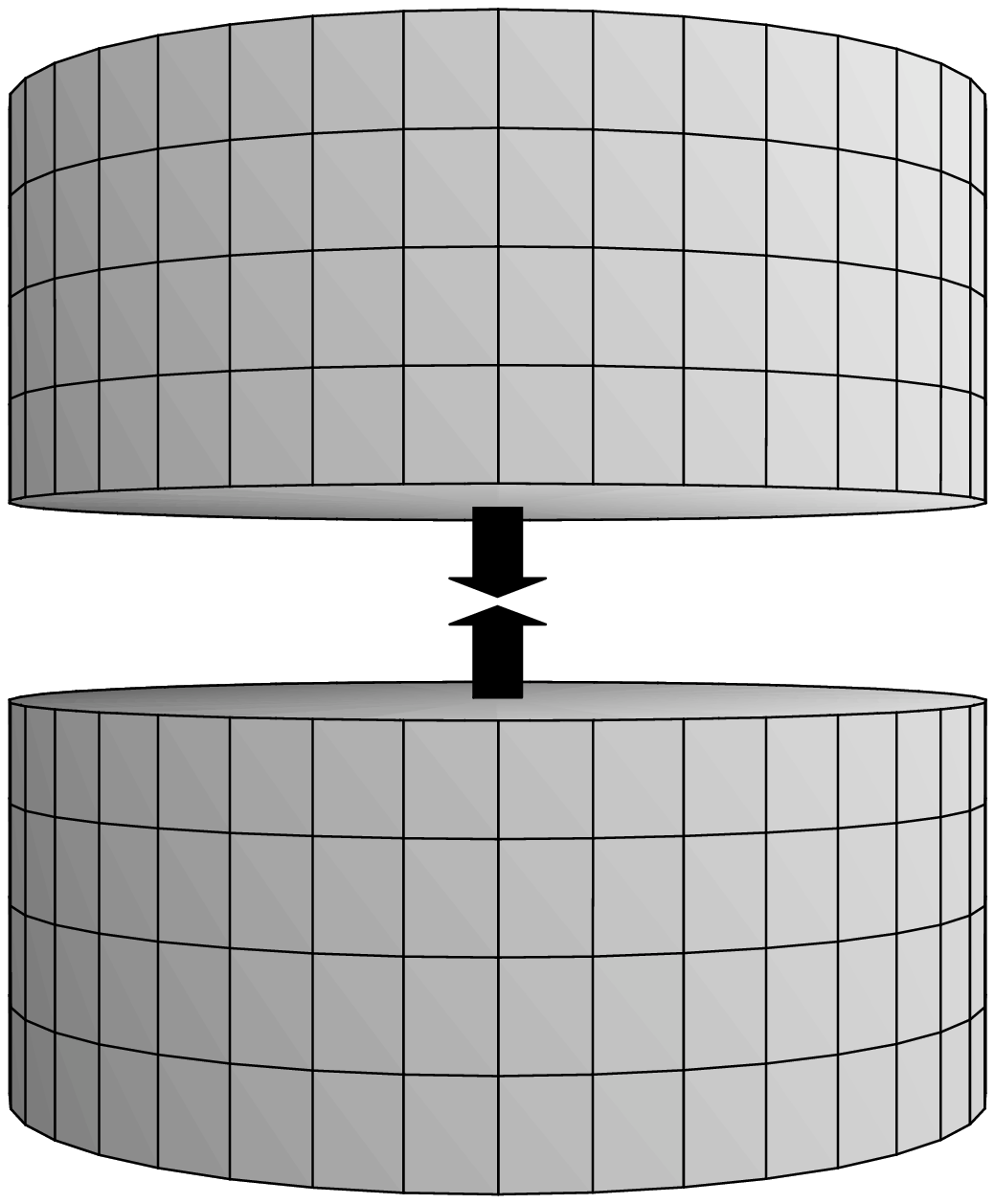}
\end{tabular}
&
\begin{tabular}{c}
\includegraphics[width=1in,height=0.375in]{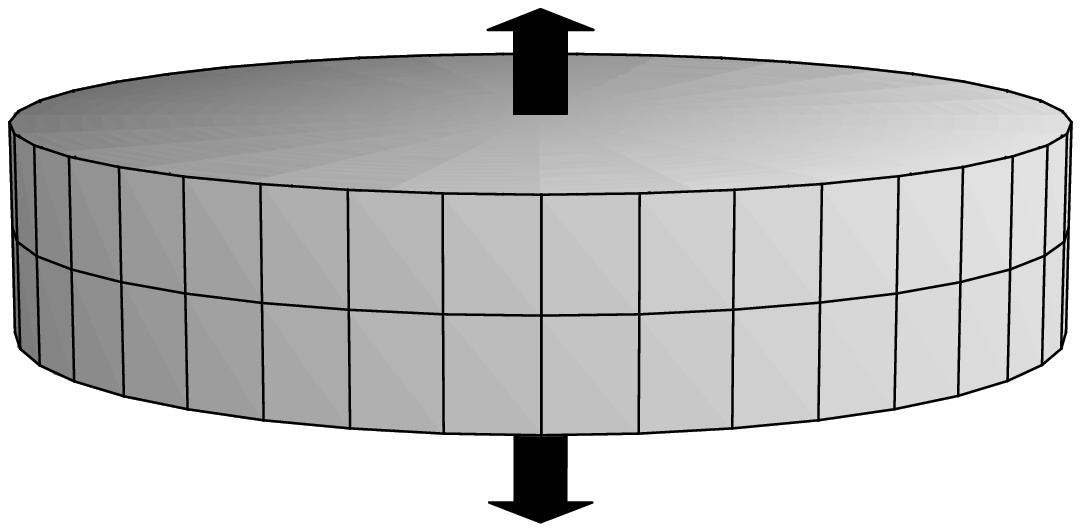}
\end{tabular}
\end{tabular}
\end{center}
\caption{
\label{fig:cylinders}
	Normal vectors used to calculate flux
}
\end{figure}

Having received this challenge, we calculated the surface mass that 
was said to be present in the four galaxies that we had studied by 
integrating (\ref{Eq2r}) over the surface without paying heed to the 
actual sign of the result. Suspicions were aroused from the discovery 
that (\ref{Eq2r}) in each case gave a numerical value slightly less 
than the mass that we had derived from the volume integral of our 
\textit{continuous} mass density distribution using (\ref{Eq14}), 
(\ref{Eq11}) and (\ref{Eq9b}).  \footnote{
	It should be noted that the two terms in (\ref{Eq2r}) were
	found to contribute equally.
}
This pointed to a plausible explanation: in our case, \textit{with 
our choice of model}, there is no \textit{physical} mass layer 
present on the $z=0$ plane. \textit{ The surface integral of this 
singular layer is merely a mathematical construct that indirectly 
describes most of the continuously distributed mass by means of the 
Gauss divergence theorem}. To see this, consider the vector ${\bf F}$ 
defined as \footnote{
	${\bf e}_r$ and ${\bf e}_z$ are unit vectors in the $r$
	and $z$ directions.
}
\eqnn{Eq3r}{
	{\bf F} \equiv A(r,z){\bf e}_r + B(r,z){\bf e}_z
}
where 
\eqnn{Eq4r}{
	(8\pi G/c^2)B \equiv \frac{NN_z}{2r^2} +\frac{N_rN_z}{2r}
}
as a first option. We choose $A(r,z)$ so that 
\eqnn{Eq5ra}{
	\int\nabla\cdot {\bf F}dV \equiv (8\pi G/c^2)M
}
where $M$ is the total mass.  As a more transparent second option,
we choose
\eqnn{Eq4ar}{
	(8\pi G/c^2)B \equiv \frac{NN_z}{r^2}
} 
where we define
\eqnn{Eq5r}{
	\nabla\cdot {\bf F} \equiv (8\pi G/c^2)\rho
}
From these definitions, we deduce the form of $A(r,z)$ in order to 
produce the density as expressed through $N$ in (\ref{Eq9b}).  We 
calculate the mass over the cylindrical volume defined by $-\infty 
<z<\infty$, $0<r<r_{galaxy}$.  By the Gauss divergence theorem, the 
volume integral of $\rho$, via (\ref{Eq5r}) is equal to the integral 
of the normal component of ${\bf F}$ over the bounding surfaces.  
However, the integration must be over a continuous domain and since 
the ${\bf e}_z$ component is discontinuous over the $z=0$ plane, the 
volume integral must be split into an upper and a lower half. The two 
new surface integrals together would constitute the jump integral of 
(\ref{Eq2r}) in the first option if one were to be cavalier about the 
directions of unit \textit{outward} normals, as we shall discuss in 
what follows. The surfaces above and below the galaxy give zero 
because of the exponential factors in $z$ and the final small 
contribution comes from the cylinder wall via the $A$ function.

In our solution, the actual \textit{physical} distribution of mass is 
not in concentrated layers over bounding surfaces: the Gauss theorem 
gives the value of the \textit{distributed} mass via equivalent 
purely mathematical surface constructs as we are familiar from 
elementary applications of this theorem. Physically, the density is 
well defined and continuous throughout, except on the $z=0$ plane. In 
fact the limits as $z =0$ is approached give the same finite values 
from above and below. While the field equations break down at $z=0$, 
the density for a physically viable model is logically defined by 
this limit at $z=0$. However, with the chosen form of solution, the 
density \textit{gradient} in the $z$ direction is discontinuous on 
the $z=0$ plane. This gradient undergoes a reversal for a galactic 
distribution with diminishing density in both directions away from 
the symmetry plane. It is most convenient to achieve this with an 
abrupt reversal as we have done.  There is no indication that this 
choice alters the essential physics.

Thus we have shown via the Gauss divergence theorem, that the 
supposed surface layer is merely a re-expression of the integrals 
that constitute the \textit{continuous volume distribution} of mass.  
Indeed if one were to reject this interpretation and insist that 
these surface integrals reveal additional mass in the form of a 
layer, then the Gauss theorem would indicate that this mass must be 
negative. Indeed various authors (e.g. \cite{VL} \cite{Bonnor2}) have 
referred to negative mass layers. However, as Bondi had emphasized in 
his writings, negative mass repels rather than attracts.  Therefore 
we had set out to test the viability of the presence of such negative 
mass to see if repulsion rather than attraction was in evidence. We 
considered a test particle in our model that was comoving with the 
rotating dust apart from having a component of velocity $U^z$ normal 
to the $z=0$ plane.  The geodesic equation in the $z$ direction 
reduces to  
\eqnn{Eq50r}{
    \frac{dU^z}{ds}= \frac{N_rN_z(U^z)^2}{2r}
}
We had computed the complete $N$ series for the galaxy NGC7331 (see 
\cite{CT}). We then focused upon points in the range $r=0.1$ to $30$ 
and points above the $z=0$ symmetry plane $z=0.001$ to $1$ for the 
right hand side of (\ref{Eq50r}). All of the points gave a negative 
value as expected for the $z$ acceleration (i.e. attraction) of a 
particle in the region above the symmetry plane.  However, if the 
$z=0$ surface actually harboured a \textit{physical} negative mass 
surface layer, indeed one of numerical value comparable to the 
positive mass of the normal galactic distribution, 
\footnote{
In an interesting recent paper \cite{Bonnor2}, the motion of a test particle for a different type of distribution was analyzed in the locally non-rotating frame produced via the transformation in (\ref{Eq3}). In this case, the $z$ geodesic equation is dominated by  $\Gamma^z_{00}$ for non-relativistic particles. In \cite{Bonnor2}, this term indicated that $dU^z/ds$ was positive for $z>0$ and hence implied an apparent repulsion of the test particle. However, the geodesic equation should apply to particles of the dust itself since they are geodesic. These particles are at rest in the original frame and have only tangential velocity in the locally non-rotating frame. The local transformation should not alter their having no $z$ velocity yet there is an apparent $z$ acceleration. 
This is a contradiction which is resolved as follows: while we can use the local transformation to derive the local angular velocity (and hence tangential velocity) of the particles, it is not legitimate to take derivatives of the metric found after the local transformation has been applied in order to derive acceleration. The former usage simply reads off the required angular velocity to diagonalize the metric locally. No differentiation is required to do so. However, the latter usage would be legitimate only if the transformation would have been effected \textit{without} constraints, in this case the constraint of holding $r$ and $z$ fixed. By holding these fixed to derive the new $ds^2$, we have metric tensor components that are correct as such only through a different transformation at each point. Hence there is an inherent discontinuity of transformation. In this manner, we recognize the derivative required to find $dU^z/ds$ as being illegitimately applied, thus resolving the contradiction. 
This also brings into question the interpretation of the lack of a $1/r$ term in $g_{00}$ in the locally non-rotating frame for the same reason. The $1/r$ issue is a global one yet the transformation that brought the metric to the form that is being used is a purely local one. By contrast, no such problem arises in studying test particle motion relative to the dust co-moving frame with the particle following the motion of the dust cloud apart from a $z$ velocity component. 
The result is logical: for $U^r$ and $U^{\phi}$ being zero, acceleration occurs only for $U^z$ different from zero (otherwise it would be part of the dust cloud and hence stationary) and the acceleration is independent of the sign of the velocity. Moreover, the acceleration is \textit{negative} for $z>0$ and \textit{positive} for $z<0$. Thus, it is \textit{attracted} to the central plane in both cases.
We thank Professor W.B. Bonnor for bringing his paper to our attention.
}
then at the very 
least, one would have expected to witness a  \textit{repulsion} of 
the particle as the test particle approached the boundary. The 
absence of this occurence adds further support to our original model 
\cite{CT} as being free of surface layers of mass.

It is true that our choice of solution leads to a discontinuity in 
the z-derivative of $N$ across the $z=0$ plane. This goes hand-in-hand 
with the physically natural density \textit{gradient} discontinuity 
across the symmetry plane. \footnote{
	This is even more benign than the density discontinuity
	in the constant density Schwarzschild sphere solution.
} 
To see this, consider the essential characteristics of our model 
which consists of dust with reflection symmetry about the $z=0$ 
plane. The density naturally increases symmetrically as this plane is 
approached from above and from below with the same absolute value but 
opposite sign from symmetry. In all generality, the density gradient 
will be different from zero as this plane is approached and  because 
of reflection symmetry, the gradient will of necessity be 
discontinuous \footnote{
	The density gradient is governed by the behavior of odd 
	derivatives of $N$ with respect to $z$. However, the 
	density itself is governed by $N_z^2$ (\ref{Eq9b}) which 
	has the same limit as z approaches $0$ from above or below. 
	Thus, we define the value of $\rho$ at $z=0$ by this common 
	limit and hence the singularity is removable.
}. It is only with delicate fine-tuning that this discontinuity can 
be avoided and this will be the case only if the density gradient is 
adjusted to be precisely zero as the $z=0$ plane is approached from 
above and below. 

\begin{figure}
\begin{center}
\includegraphics[width=2.75in]{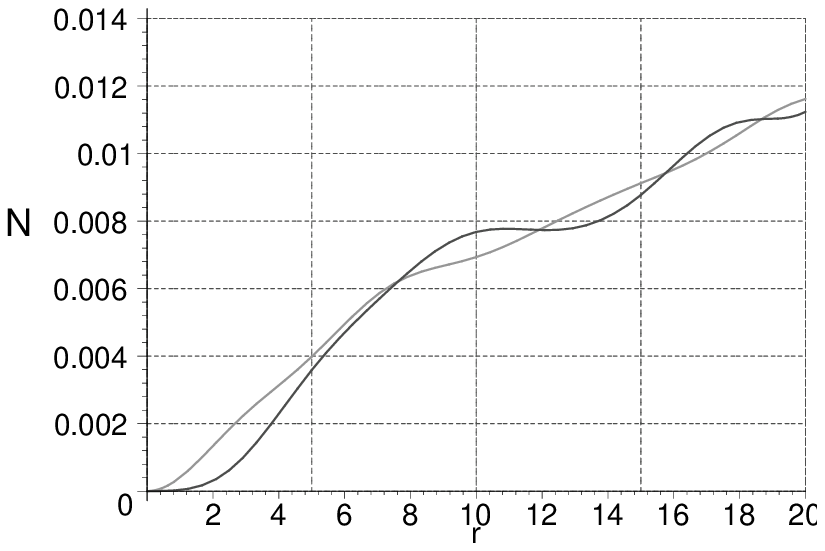}
\includegraphics[width=2.75in]{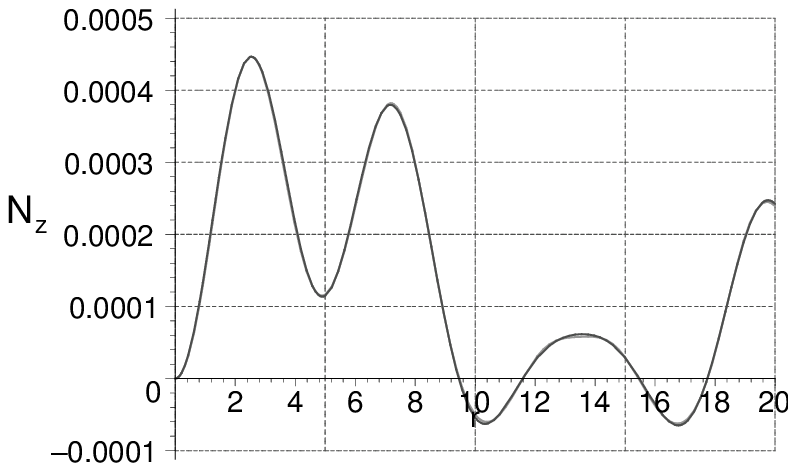}
\end{center}
\caption{
\label{fig:matching}
	Matching conditions for $N$ and $\partial N/\partial z$
	at $z=z_0$.
}
\end{figure}

As an exercise in response to critical comments \cite{CT2}, we  achieved 
this approximately by choosing $\cosh(\kappa_n z)$ functions in place 
of exponential functions to span the region in a sandwich 
encompassing the symmetry plane and employing the usual exponential 
functions beyond this sandwich. This led to the issue of matching the 
$N$ and $N_z$ functions along the external/internal region joins and 
it was achieved by using many different $k_n$ parameters for the 
external exponential functions as opposed to the original 10 internal 
parameters of the original model.  In \cite{VL}, it was claimed that 
a matching could not be achieved but these authors had not realized 
that we used different and many parameters for the outside regions. 
Since then, we have refined the fit further by employing hundreds of 
external parameters and the improved fit is shown in Figure 
\ref{fig:matching}. However, it must be stressed that the generic 
situation would be one in which the density gradient is discontinuous 
at $z=0$.

In \cite{garf}, the well-known expression of the field equations in 
the harmonic gauge in Cartesian coordinates
\eqnn{Eq1g}{
     \partial_k\partial^kh^{ab}= \frac{16{\pi}G}{c^4}\tau^{ab}
}
($\tau^{ab}$ includes the energy-momentum tensor of the matter plus 
the non-linear terms in the Einstein equations) is invoked. (A 
related line of reasoning was followed in \cite{korz}). In 
\cite{garf}, the author presents the standard description of the 
post-Newtonian perturbation scheme to conclude that the solution to 
the galactic problem must be the usual Newtonian one and that all 
corrections must be of higher order.  Firstly, we did not use this 
scheme (as noted as well in \cite{maia}).  Just as one would not 
logically choose Cartesian coordinates in the harmonic gauge to 
describe FRW cosmologies, one would not normally choose these for our 
stationary axially symmetric galactic problem. Our problem is greatly 
simplified with cylindrical polar coordinates co-moving with the 
matter.  Secondly, for the gravitationally bound system under study, 
the metric components are of \textit{different} orders in $G$ 
\footnote{
	In general, first order perturbations lead to linear 
	equations. However, there are situations where this is not 
	the case. For example in fluid mechanics, certain 
	approximations still lead to non-linear equations as in 
	Burger's equation.
}.
If one were to take the approach suggested in \cite{garf}, the 
equations (\ref{Eq1g}) could be schematically expressed as
\eqnn{Eq2g}{
	\nabla^2h_{(1/2)}=0,
	\quad \nabla^2h_{(1)} = GT  + h_{(1/2)}^2
}
where tensorial superscripts have been suppressed and the lower case 
numbers refer to orders in $G$. In this manner, we would have 
incorporated the non-linear structure of our system within the 
framework of the scheme suggested by \cite{garf}. The novel aspect is 
that the lowest order equation (of order $G^{1/2}$) in (\ref{Eq2g}) 
has zero on the RHS and the second equation that would normally be 
the Newtonian Poisson equation, differs in that it has non-linear 
terms.  Thus, the structure of our solution does not proceed as in 
the standard approach of (\ref{Eq1g}). In the latter standard 
approach, the lowest order base solution is the Newtonian solution 
whereas in the galactic problem, the lowest order equation is the 
Laplace equation for which an order $G^{1/2}$ solution is necessary 
(see \cite{C} where this component is inappropriately chosen to be 
zero) and the next order (order $G^1$) equation for the density 
(\ref{Eq3g}), 
\eqnn{Eq3g}{
	\frac{N_r^2 + N_z^2}{r^2} = \frac{8{\pi}G\rho}{c^2}
}
has non-linear terms in the metric in the form of the squares of the 
derivatives of an order $G^{1/2}$ metric tensor component $N$. Thus, 
our situation is unlike standard iterative perturbation scheme 
applications as envisaged in \cite{garf}. Hence there is no basis to 
draw the conclusions that are expressed therein.

Further in \cite{garf}, the author refers to ``extra matter" in the symmetry 
plane of the galaxy and muses whether our model ``could be somehow 
fixed".  However, in \cite{CT2} we presented the evidence that our 
solution embodies the physically natural density gradient 
discontinuity at the plane of symmetry and that it does not contain 
extra matter. Moreover, we showed that if there were to be a surface 
layer of mass, it would be negative mass but this was negated by the 
attraction rather than repulsion of test particles near the symmetry 
plane, as we discussed above.

The author concludes with an argument to attempt to provide dark 
matter through general relativity in the form of a geon where general 
relativity would be required and he deduces that this is impossible 
\cite{garf}. While we are in agreement with him that it is indeed 
impossible with geons (but from a different line of reasoning, see 
\cite{CFP},\cite{PC}), the argument is irrelevant because the 
galactic field is weak and hence geons are \textit{a priori} out of 
the question, even if they were viable in principle.

With regard to the issue of gauge, it was argued in \cite{korz} that 
asymptotically flat solutions are unattainable with a lead-off 
$G^{1/2}$ order metric component.  However, we have shown that they 
are readily attainable in conjunction with the physically desirable 
$N_z$ discontinuity and are approximately attainable with the 
smoothed fine-tuned solution discussed above. Moreover, they are 
precisely attainable when an essential singularity  
is invoked\footnote{
	This was almost achieved in \cite{BG}. Their axis
	singularity prevented global asymptotic flatness. However,
	exact solutions with compactified singularities of the Weyl
	type are likely to rectify this deficiency.
}.  A key point is that the equations have an inherent non-linearity 
as a result of the fact that the metric components are of different 
orders and the different orders are a necessary consequence of the 
problem being a gravitationally bound one.

In \cite{C}, the author brings up the covariant vorticity and 
(vanishing) shear tensors for the rotating dust and poses the latter  
characteristic as being inconsistent with a galaxy that has
differential rotation. However, a rotating dust cloud cannot 
physically rotate rigidly as does a disk of steel which has internal 
stresses. The answer to \cite{C} is that the vanishing 
\textit{covariant} shear is analogous to the vanishing 
\textit{covariant} acceleration of a freely moving particle. In the 
case of the latter, it is only under very special conditions that the 
motion will be one of constant velocity. The generic motion will be 
conical or more complicated. This could have been recognized in 
\cite{C} where the correct \textit{differential} local angular 
velocity $cN(r,z)/r^2$  is 
displayed. Also in \cite{C}, when the author chooses a solution for 
which the $N$ function is taken to be zero at order $G^{1/2}$, he is 
being inconsistent with the demand that this is a gravitationally 
bound problem with rotation. Finally this author treats the 
transformation $ \phi \rightarrow \bar\phi = \phi + \omega(r,z)t$ as a 
`global' transformation to the `co-moving frame'. However it is the 
original coordinate set that constitutes the co-moving frame and 
moreover, this transformation has value strictly as a local 
transformation.

The authors of \cite{BG} arrive at our equations (\ref{Eq5top}),
(\ref{Eq5})
(with $w$ set to zero) apart from the exponential $\nu$ factor which 
they later note can be taken to be a constant scaling factor and find 
the same order of magnitude reduction of galactic mass that we had 
found \cite{CT} starting from their exact solution class.  This 
provides some vindication for our analysis. It should be noted that 
their scaling factor is actually incorporated in our solutions within 
the computed amplitudes of our basis expansion functions. To be 
particularly noted  in \cite{BG} is that their solution class is 
fine-tuned as the density gradient is precisely zero at $z=0$. The 
price that is paid to achieve this degree of smoothness is the 
incorporation of an axial singularity.  These authors justify the 
singularity by identifying it as a jet.  While jets are observed in 
various galaxies in their formative stages, they are not known to be 
present in the essentially stationary galaxies that are being modeled 
with this class.

A detailed analysis of the exact van Stockum \cite{vs} spacetime is 
provided in \cite{BJK} from which the authors derive and transfer 
supposed restrictions onto our work. However in doing so, they miss 
the point that we are analyzing in generality the \textit{weak-field} 
stationary axially symmetric dust spacetimes and hence restrictions 
derived on an exactness basis from some particular solution are not 
relevant to us. Our solutions are approximate, with sufficient 
accuracy for the physical situation at hand. 

\begin{figure}
\begin{center}
\includegraphics[width=3in]{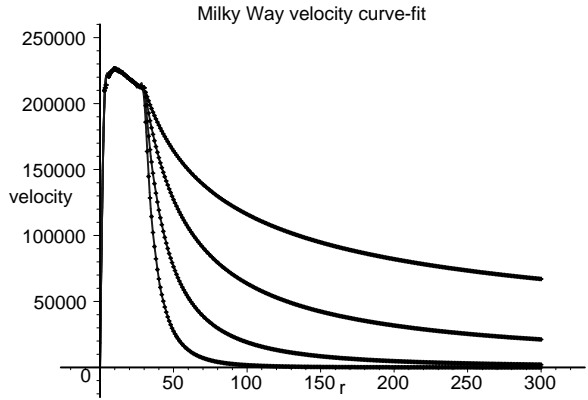}
\end{center}
\caption{  
\label{fig:milkywayvelcfalloff}
	Beyond the HI region, the velocity can be modeled in many 
	different manners: here $V\propto1/\sqrt{r}$,  
	$V\propto1/r$, $V\propto1/r^2$ and $V\propto1/r^4$ are 
	illustrated.
}
\end{figure}

Moreover, we have now considered various velocity fall-off scenarios 
beyond the HI regions and have extended our rotation curves to match 
these assumed fall-offs.\footnote{
	It is to be noted that this is in keeping with our desire 
	to work with globally dust models, thus avoiding transition 
	issues for the metric and its derivatives in going from 
	dust to total vacuum. Also to be stressed is that the 
	points beyond the HI region are not based upon observed 
	data but rather are artificially imposed to induce smooth 
	matter fall-offs of various forms.
} 
These are shown in figure \ref{fig:milkywayvelcfalloff}. To 
accommodate this expanded region, this requires a completely new and 
enlarged set of parameters from those displayed in the Tables to 
follow the relatively flat region and then merge into the fall-off 
region. It is to be noted that since the velocity continues to be 
constructed in the form of (\ref{Eq14}), there is continuity of the 
curve and its derivatives apart from the value at precisely $z=0$ 
discussed previously. The kink in figure 
\ref{fig:milkywayvelcfalloff} is only apparent, arising from the 
practical need to fit the subtle transition into a compressed graph.

\begin{figure}
\begin{center}
\includegraphics[width=3in]{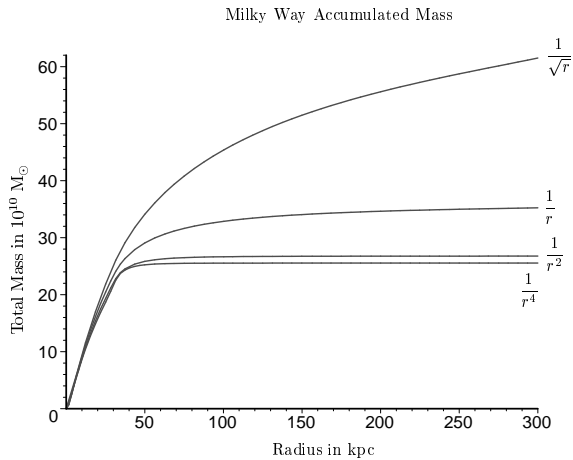}
\end{center}
\caption{  
\label{fig:milkywayaccmass}
        The Milky Way's accumulated mass as a consequence of
        velocity faall-off beyond the HI region.	
}
\end{figure}

From (\ref{Eq9b}), we see that the density vanishes when $N_r$ and 
$N_z$ are both zero, i.e. when $N$ is a constant. Also, from 
(\ref{Eq11}), when $N$ is a constant, the velocity falls as $1/r$. 
Therefore, at first glance one might believe that by choosing the 
continued velocity curves beyond the HI region in the form $A_0/r$, 
one would be tracking $r$ into the vacuum, identifying $N$ with a 
constant $A_0$ and hence accumulating no further mass. However, while 
in plotting rotation curves at a given $z$,  
it is only a net independence in $r$ for the $N$ function that is 
required to give an $A_0/r$ fall-off in $V$. The $z$ dependence in 
the $N$ function can still be present.  Thus, the density will still 
not be zero and mass will still be accumulated as one tracks in the 
radial direction.\footnote{
	This is evident in Figure \ref{fig:milkywayaccmass}.
}  
Faster fall-off rates with $r$ would improve the trend towards vacuum 
further. Slower fall-offs such as of the form $1/\sqrt{r}$ which 
might be suggested from Newtonian gravity are clearly not adequate to 
merge towards vanishing density. \footnote{
	A fall-off of the form $1/\sqrt{r}$ would be appropriate to 
	impose for test particles in the field of a massive body 
	such as is the case in the solar system. However, here we 
	have seen that in the case of a continuous gravitationally 
	bound source, general relativity presents a dynamical 
	system with behavior that does not match the Newtonian 
	picture.  
}
It is also not necessary since we are basing our analysis on the 
preferred theory of gravity, namely general relativity. The challenge 
is to find a general relativistic solution that merges properly into 
near vacuum and we have met that challenge. 

We display the accumulated mass for the Milky Way in a highly 
extended cylindrical volume of 300 Kpc in size in figure 
\ref{fig:milkywayaccmass}. It is to be noted that even assuming a 
Newtonian-like fall-off of the form $1/\sqrt{r}$, there is a far less 
amount of accumulated mass up to a radius of ten times the visible 
radius than is envisaged by the use of Newtonian as opposed to 
general relativistic galactic dynamics. An even slower accumulation 
of mass is seen for the $1/r$ fall-off. 
For such a fall-off, the accumulated mass is approximately $35 \times 
10^{10}M_\odot$ at a radius of 300 Kpc and a linear extrapolation to 
$r$=900 Kpc yields a value of $39.2 \times 10^{10}M_\odot$, a very 
modest increase in comparison to Newtonian modeling. Moreover, the 
faster fall-offs of $1/r^2$ and $1/r^4$ yield very minor mass 
increases out to very large radii as can be seen in figure 
\ref{fig:milkywayaccmass}. \footnote{
	Note from this figure that the accumulated mass at 30 Kpc 
	is approximately the same  for the various fall-off 
	scenarios as well as the value stated in Section 3 where we 
	used only 10 parameters and where we did not focus on the 
	bahavior of the model beyond the 30 Kpc edge of the HI 
	region.
}

This fortifies our contention that general relativity obviates the need 
for overwhelmingly dominant massive halos of exotic dark matter.

It is to be noted that our models are \textit{globally} dust 
\footnote{
	We make this choice for the composition and distribution
	of the matter for the purpose of mathematical simplicity.
}
and therefore there is no basis for a matching with the vacuum Kerr 
metric asymptotically. Our models are asymptotically flat with a 
well-defined mass. The Tolman integral dictates the value of this 
mass and since there is no stress and the fields are weak, the Tolman 
mass to lowest order is simply given by the coordinate volume 
integral of the density.

In \cite{FP}, the observed solar neighborhood density data are 
compared with the values given by our model. These authors find that 
our density is less by a factor of approximately six. Firstly, it is 
to be noted that there are considerable error bars on their data and 
it is possible that our neighborhood could be one of enhanced 
density. Secondly,  it must be emphasized that our model is adjusted 
in the simplest terms to account for the overall mean velocity 
distribution with a mere ten parameters. \footnote{
	Even with the enlarged parameter set used to model the
	asymptotic fall-off, the parameter input is still quite modest.
}
This scant input can hardly be expected to account for the 
distribution of hundreds of billions of stars and their motions. This 
is beyond the capacity of even Newtonian theory let alone general 
relativity. By necessity, great simplifications are necessary in 
practice. Moreover, as additional support for our model having the 
correct overall characteristics, our integrated density over the 
visible region falls within predicted limits.

Even if one were to assume a logical basis for making their 
comparison between local observed data and our globally derived very 
approximate data, it should be noted that the vertical 
distribution of observed stars in the local galactic disk has a sharp 
peak \cite{HF}. This fits in well with the presence of a density 
gradient discontinuity in our models. Moreover, it should be noted 
that for mathematical tractability, we have assumed that the density 
peaks all occur at $z=0$. However, there will naturally be some 
variation in the location in $z$ for these peaks. Due to the rapid 
decline in density, there will be large local variations in density 
and therefore the criticisms in \cite{FP} about our distributions 
are further seen to be inappropriate.

In \cite{K}, the author models the galaxy via Newtonian physics with 
a surface mass layer. Clearly, given the freedom to impose the 
properly adjusted internal stresses, virtually any kind of 
approximate velocity distribution can be simulated. This is 
inadequate for the problem at hand on two counts: firstly, a good 
model should be stress-free (i.e. purely free-fall gravitationally 
driven) and secondly, the point is to develop an extended 
\textit{volume} distribution and hence more akin to reality. This is 
what we have set out to do and this, within general relativity, the 
preferred theory of gravity. Moreover, there is the suggestion, 
sometimes reiterated by others, that we claim to have modeled 
galaxies without any dark matter, this in spite of the fact that we 
have explicitly referred to dead stars, neutron stars, etc.  as dark 
matter constituents of galaxies and we have presented mass-to-light 
ratios. What we are questioning is the existence of \textit{exotic} 
dark matter, the supposed constituent of the massive halos 5 to 10 
times the traditionally computed mass that are said to surround 
galaxies, matter that has no known counterpart to the matter that 
physicists identify in labs and particle accelerators. Our approach 
is in keeping with the spirit of Occam.

In \cite{MM}, the authors fault our models as extended constructs 
that indicate enormous quantities of mass beyond the HI regions.  
This is a useful point of criticism in that we had not investigated 
earlier the asymptotic consequences of the particular parameter sets 
that we had chosen to model the observed rotation curves. \footnote{
	Note however that their argument that mass accumulates 
	linearly in $r$ is faulty as a generalization. With the 
	correct combination of parameters, the term that would lead 
	to such an accumulation can be eliminated. Our examples in 
	which we achieve minimal accumulation, provide the direct 
	proof that this is the case.
} 
However, in this paper, as first reported in \cite{CT3}, we assure 
more realistic fall-off scenarios \footnote{
	It is to be noted that in so doing, while the expansion 
	parameters are no longer the same as in the earlier sets, 
	we have determined that the net physical effects are of 
	insignificant difference within the observed matter 
	distribution in the two approaches.
}
and we find that the  accumulated mass profiles indicate that most of 
the mass of a galaxy is confined fairly close to the region of the 
visible disk with \textit{modest} accumulations of mass beyond this 
region. General relativity achieves this with a pressure-free fluid 
model, unlike Newtonian gravity. 

The comments above referring to \cite{BJK} apply also to \cite{ZAT}. 
Moreover, as we have reported at various seminars in June/July 2006 
and as described above, we have avoided the complications of merging 
from the dust regions to vacuum by dealing with models that are 
globally dust. Since the dust in our models become extremely diffuse 
with distance, the physical distinction between having the global 
dust and the vacuum is inconsequential. Also to be noted is that 
since the $N$ function is ultimately connected to solutions of the 
Laplace equation, there will necessarily be a singularity of some 
form present. It is quite acceptable if it is the right kind of 
singularity and in our construction it is the case, modeling the 
physically desirable density gradient discontinuity.

In an interesting approach from a very different direction, Lusanna 
\cite{Lus} has pointed to relativistic inertial effects that do not 
have a Newtonian limit counterpart. He has suggested that in the weak 
field limit, these effects could match our results. 

\section{Velocity Dispersion Test}

\begin{figure}
\begin{center}
\includegraphics[width=3in]{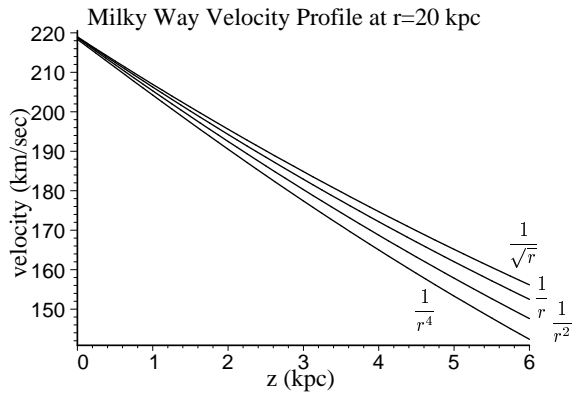}
\end{center}
\caption{
\label{fig:milkywayvdispersion}
	Velocity dispersion at $r=20$ kpc for the Milky Way.
}
\end{figure}

Clearly it is important to approach the exotic dark matter issue in 
as many ways as possible. After all, from a purely formal point of 
view, general relativity should be able to model vastly extended 
distributions of pressure-free fluids in rotation.  In this vein, we 
have constructed a test in principle that relies upon data in the 
\textit{visible/HI} regime thus making it particularly useful. When we examine Figure \ref{fig:milkywayaccmass},
we see that different constructed velocity fall-off profiles beyond
the HI region imply different mass accumulations in those external
regions. Carrying these back with continuity into the visible/HI
region, we find that the extent of the velocity dispersion as we
track curves at different non-zero $z$ values depends on the assumed
external velocity profile fall-off. (See, for example.
Figure \ref{fig:milkywayvdispersion}.)

Thus, it is our request to our astronomical colleagues to kindly 
provide us with the data for rotation curves in planes of different 
$z$ values. With sufficient data, it should be possible, at least in 
principle, to provide limits on the extent of extra matter that might 
lie outside of the visible/HI region.

\section{Concluding Comments}

One might be inclined to question how this large departure from the 
Newtonian picture regarding galactic rotation curves could have 
arisen since the planetary motion problem is also a gravitationally 
bound system and the deviations there using general relativity are 
so small. The reason is that the two problems are very different: in 
the planetary problem, the source of gravity is the sun and the 
planets are treated as test particles in this field (apart from
contributing minor perturbations when necessary). They respond to 
the field of the sun but they do \textit{not} contribute to the 
field. By contrast, in the galaxy problem, the source of the field 
is the combined rotating mass of all of the freely-gravitating 
elements themselves that compose the galaxy.

We have seen that the non-linearity for the computation of density 
inherent in the Einstein field equations for a stationary 
axially-symmetric pressure-free mass distribution, even in the case 
of weak fields, leads to the correct galactic velocity curves as 
opposed to the incorrect curves that had been derived on the basis of 
Newtonian gravitational theory. Indeed the results were consistent 
with the observations of velocity as a function of radius plotted as 
a rise followed by an essentially flat extended region and no halo of 
exotic dark matter with multiples of the normally computed galactic 
mass was required to achieve them. The density distribution that is 
revealed thereby is one of an 
essentially flattened disk without an accompanying overwhelmingly
massive vastly extended dark matter halo. With the ``dark'' matter 
being associated with the disk which is itself visible, it is
natural to regard the non-luminous material as normal baryonic
matter.

It is unknown how far the galactic disks extend. More data points 
beyond those provided thus far by observational astronomers would 
enable us to extend the velocity curves further. We have made 
simplifying assumptions for various velocity fall-off scenarios and 
we have seen that these can readily yield a picture of galactic 
structure devoid of huge extended very massive halos of exotic dark 
matter.

Of particular interest is that we have within our grasp a criterion 
for determining the extent, if of any significance, of extra matter 
beyond the visible and HI regions of a galaxy. It is possible in 
principle to determine this with data solely \textit{within} the 
visible/HI region by plotting the velocity dispersion of rotation 
curves for various $z$ values. This is an attractive area for future 
research.

Nature is merciful in providing one linear equation that enables us 
by superposition to model disks of variable density distributions.  
This opens the way to studies of other sources and with further 
refinements. It is to be emphasized that what we have taken is a 
first step, a general relativistic as opposed to a Newtonian analysis 
at the galactic scale. It will be of interest to extend this general 
relativistic approach, with the hitherto neglected consideration of 
non-linearities, to the other relevant areas of astrophysics with the 
aim of determining whether there is any scope remaining for the 
presence of any exotic dark matter in the universe. For example, at 
the scale of clusters of galaxies, the virial theorem of Newtonian 
physics is used. However, such a system, albeit now chaotic, can 
again be viewed as a continuum of free-fall matter as was the case 
for the galactic scale. 
Indeed at the scale of individual galaxies as units within the cluster, the motions comprise a multitude of randomly oriented free-fall-rotations . While the chaotic nature of these rotations within a cluster might have the effect of minimizing or even erasing the kind of phenomenon that we have witnessed in the systematic rotation of an individual galaxy, it might be otherwise.
Since general relativity was seen to make 
such a difference in the case of the galactic scale, clearly it is necessary 
to analyze the scale of the clusters anew.  

The scientific method has been most successful when directed by 
Occam's razor, that new elements should not be introduced into a 
theory unless absolutely necessary. If it should turn out to be the 
case that the observations of astronomy can ultimately be explained 
without the addition of new exotic dark matter, this would be of 
considerable significance.

\acknowledgments

We thank our many colleagues for their interest in our work, their 
advice and criticism. This work was supported in part by a grant from 
the Natural Sciences and Engineering Research Council of Canada.

\appendix
\section{Appendix}

The coefficients for
\eqn{
	N(r,z)= -\sum_{n=1}^{10} C_n k_n r e^{-k_n |z|}J_1(k_nr)
}
are tabulated in Tables \ref{table:milkyway} to \ref{table:ngc7331}
with $r$ and $z$ in Kpc. The velocity in m/sec is given by
\eqn{
	V(r,z)= \frac{3\times 10^8}{r} N(r,z)
}
and the density in kg/m$^3$ is given by
\eqn{
	\rho(r,z) = 5.64\times 10^{-14}
	\frac{ \left( N_r^2 + N_z^2 \right) }{r^2}
}

\begin{table}[ht]
\begin{center}
\begin{tabular}{c c}
\hline
$-C_n k_n$ & $k_n$ \\
\hline
0.0012636497740 & 0.06870930165 \\
0.0004520156256 & 0.15771651740 \\
0.0001785404942 & 0.24724936890 \\
0.0002946610499 & 0.33690098400 \\
0.0000103378815 & 0.42659764880 \\
0.0002127633340 & 0.51631611340 \\
-0.0000221015927 & 0.60604676080 \\
0.0001346275993 & 0.69578490080 \\
-0.0000123824930 & 0.78552797510 \\
0.0000666973093 & 0.87527447050 \\
\hline
\end{tabular}
\caption{
	\label{table:milkyway}
	Curve-fitted coefficients for the Milky Way
}
\end{center}
\end{table}

\begin{table}[ht]
\begin{center}
\begin{tabular}{c c}
\hline
$-C_n k_n$ & $k_n$ \\
\hline
0.0011694103480 & 0.1093102526 \\
0.0004356556836 & 0.2509126413 \\
0.0003677376760 & 0.3933512687 \\
0.0001484103801 & 0.5359788381 \\
0.0000837048346 & 0.6786780777 \\
0.0000414084713 & 0.8214119986 \\
0.0000429277032 & 0.9641653013 \\
0.0000550130755 & 1.1069305240 \\
0.0000238560073 & 1.2497035970 \\
0.0000129841761 & 1.3924821120 \\
\hline
\end{tabular}
\caption{
	\label{table:ngc3031}
	Curve-fitted coefficients for NGC 3031
}
\end{center}
\end{table}


\begin{table}[ht]
\begin{center}
\begin{tabular}{c c}
\hline
$-C_n k_n$ & $k_n$ \\
\hline
0.00093352334660 & 0.07515079869 \\
0.00020761839560 & 0.17250244090 \\
0.00022878035710 & 0.27042899730 \\
0.00009325578799 & 0.3684854512 \\
0.00007945062639 & 0.4665911784 \\
0.00006081834319 & 0.5647207491 \\
0.00003242780880 & 0.6628636447 \\
0.00003006457058 & 0.7610147353 \\
0.00001687931928 & 0.8591712228 \\
0.00003651365250 & 0.9573314522 \\
\hline
\end{tabular}
\caption{
\label{table:ngc3198}
	Curve-fitted coefficients for NGC 3198
}
\end{center}
\end{table}


\begin{table}[ht]
\begin{center}
\begin{tabular}{c c}
\hline
$-C_n k_n$ & $k_n$ \\
\hline
0.0015071991080 & 0.0586542819 \\
0.0003090462519 & 0.1346360514 \\
0.0003960391396 & 0.2110665344 \\
0.0001912008955 & 0.2875984009 \\
0.0002161444650 & 0.3641687246 \\
0.0000988404542 & 0.4407576578 \\
0.0001046496277 & 0.5173569909 \\
0.0000619051218 & 0.5939627202 \\
0.0000647087250 & 0.6705726616 \\
0.0000457420923 & 0.7471855236 \\
\hline
\end{tabular}
\caption{
\label{table:ngc7331}
	Curve-fitted coefficients for NGC 7331
}
\end{center}
\end{table}


\end{document}